\documentclass[aps,prd,superscriptaddress,showpacs,nofootinbib,
floatfix]{revtex4}
\usepackage{epsfig}

\def\be{\begin{equation}}
\def\ee{\end{equation}}
\def\bea{\begin{eqnarray}}
\def\eea{\end{eqnarray}}

\def\reff#1{\ref{#1}}
\def\labell#1{\label{#1}}

\def\app#1{Appendix~\reff{#1}}
\def\apps#1#2{Appendices~\reff{#1}--\reff{#2}}
\def\eq#1{Eq.~(\reff{#1})}
\def\eqs#1#2{Eqs.~(\reff{#1})--(\reff{#2})}
\def\fig#1{Fig.~\reff{#1}}
\def\figs#1#2{Figs.~\reff{#1}--\reff{#2}}
\def\sec#1{Sec.~\reff{#1}}
\def\tab#1{Table~\reff{#1}}
\def\tabs#1#2{Tabs.~\reff{#1}--\reff{#2}}

\def\cB{\mathcal{B}}
\def\cO{\mathcal{O}}
\def\cP{\mathcal{P}}
\def\cR{\mathcal{R}}
\def\ra{\rangle}
\def\la{\langle}

\def\l{\left}
\def\r{\right}

\def\Tr{\,\mathrm{Tr}}
\def\mev{\mathrm{Me\kern-0.1em V}}
\def\gev{\mathrm{Ge\kern-0.1em V}}
\def\fm{\mathrm{fm}}
\def\re{\mathrm{Re}}
\def\im{\mathrm{Im}}

\def\msbar{{\overline{\mathrm{MS}}}}
\def\NDR{\mathrm{NDR}}
\def\RI{\mathrm{RI}}
\def\RGI{\mathrm{RGI}}
\def\BSM{\mathrm{BSM}}

\newcommand{\lsim}{ {\
\lower-1.2pt\vbox{\hbox{\rlap{$<$}\lower5pt\vbox{\hbox{$\sim$}}}}\ } }
\newcommand{\gsim}{ {\
\lower-1.2pt\vbox{\hbox{\rlap{$>$}\lower5pt\vbox{\hbox{$\sim$}}}}\ } }

\begin{document}

\preprint{CPT-P07-2006}


\title{$K^0$-$\bar K^0$ mixing beyond the standard model and CP-violating
  electroweak penguins in quenched QCD with exact chiral symmetry}



\author{Ronald Babich}
\affiliation{Department of Physics, Boston University, 590
Commonwealth Avenue, Boston MA 02215, USA}
\author{Nicolas Garron}
\affiliation{DESY, Platanenallee 6, D-15738 Zeuthen, Germany}
\author{Christian Hoelbling}
\affiliation{Department of Physics, Universit\"at\,Wuppertal,
Gaussstr.\,20, D-42119\,Wuppertal, Germany}
\author{Joseph Howard}
\affiliation{Department of Physics, Boston University, 590
Commonwealth Avenue, Boston MA 02215, USA}
\author{Laurent Lellouch}
\email[]{lellouch@cpt.univ-mrs.fr}
\affiliation{Centre de Physique Th\'eorique,
CNRS Luminy, Case 907,
F-13288 Marseille Cedex 9,
France}
\thanks{CPT is ``UMR 6207 du CNRS et des universit\'es 
d'Aix-Marseille I, II et du Sud Toulon-Var, affili\'ee \`a la FRUMAM.'' }
\affiliation{Department of Physics, Universit\"at\,Wuppertal,
Gaussstr.\,20, D-42119\,Wuppertal, Germany}
\author{Claudio Rebbi}
\affiliation{Department of Physics, Boston University, 590
Commonwealth Avenue, Boston MA 02215, USA}



\begin{abstract}
We present results for the $\Delta S=2$ matrix elements which are
required to study neutral kaon mixing in the standard model (SM) and
beyond (BSM). We also provide leading chiral order results for the
matrix elements of the electroweak penguin operators which give the
dominant $\Delta I=3/2$ contribution to direct CP violation in
$K\to\pi\pi$ decays. Our calculations were performed with Neuberger
fermions on two sets of quenched Wilson gauge configurations at
inverse lattice spacings of approximately $2.2\,\gev$ and
$1.5\,\gev$. All renormalizations were implemented non-perturbatively
in the RI/MOM scheme, where we accounted for sub-leading operator
product expansion corrections and discretization errors. We find
ratios of non-SM to SM matrix elements which are roughly twice as
large as in the only other dedicated lattice study of these
amplitudes. On the other hand, our results for the electroweak penguin
matrix elements are in good agreement with two recent domain-wall
fermion calculations. As a by-product of our study, we determine the
strange quark mass. Our main results are summarized and discussed in
\sec{sec:finalres}. Within our statistics, we find no evidence for
scaling violations.

\end{abstract}



\maketitle


\section{Introduction}

Flavor changing neutral currents (FCNC), and amongst them neutral kaon
mixing, have been an extremely rich source of information about
electroweak interactions and have been highly instrumental in
designing the current standard model (SM) of particle physics. Like
all FCNC in the SM, $\Delta S=2$ processes only appear at one loop and
are highly suppressed. Therefore, they are also a sensitive probe for
new physics and provide a rich supply of constraints for models of
physics beyond the standard model (BSM), such as supersymmetric models,
left-right symmetric models, multi-Higgs models, etc.

In the SM, neutral kaon mixing is dominated by $W$-exchange box
diagrams whose short-distance contributions can be described by an
effective Hamiltonian in which the $W$-bosons and up-type quarks have
been integrated out. Because of the left-handed nature of the coupling
of $W$-bosons to quarks, only one four-quark operator
arises:
\be
O_1=[\bar s^a\gamma_\mu(1-\gamma_5) d^a]
[\bar s^b\gamma^\mu(1-\gamma_5)d^b]
\ ,
\labell{eq:smds2op}
\ee
where $a$ and $b$ are color indices.

On the other hand, in models of new physics, where heavy right-handed
particles can also experience flavor changing interactions, the
effective $\Delta S=2$ Hamiltonian is generically of the form:
\be
\mathcal{H}_\mathrm{eff}^\mathrm{\Delta
  S=2}=\sum_{i=1}^5C_i(\mu) O_i+\sum_{i=1}^3\tilde
C_i(\mu) \tilde O_i
\ ,
\labell{eq:ds2effham}
\ee
where, in addition to the SM operator of \eq{eq:smds2op},
there are the four operators, which we
give here in the supersymmetric basis of \cite{Gabbiani:1996hi}:
\bea
O_2&=&[\bar s^a(1-\gamma_5)d^a][\bar s^b(1-\gamma_5)d^b]\nonumber\\
O_3&=&[\bar s^a(1-\gamma_5)d^b][\bar s^b(1-\gamma_5)d^a]\nonumber\\
O_4&=&[\bar s^a(1-\gamma_5)d^a][\bar s^b(1+\gamma_5)d^b]
\labell{eq:susyds2ops}\\
O_5&=&[\bar s^a(1-\gamma_5)d^b][\bar s^b(1+\gamma_5)d^a]\nonumber
\ ,
\eea
as well as the parity transformed operators \cite{Gabbiani:1996hi}:
\bea
\tilde O_1&=&[\bar s^a\gamma_\mu(1+\gamma_5)d^a][\bar s^b\gamma^\mu(1+\gamma_5)d^b]\nonumber\\
\tilde O_2&=&[\bar s^a(1+\gamma_5)d^a][\bar s^b(1+\gamma_5)d^b]
\labell{eq:susyds2opstilde}\\
\tilde O_3&=&[\bar s^a(1+\gamma_5)d^b][\bar s^b(1+\gamma_5)d^a]\nonumber
\ .
\eea

In \eq{eq:ds2effham}, the Wilson coefficients $C_i$ and $\tilde C_i$
incorporate the short-distance physics associated with the new heavy
degrees of freedom which may participate in extensions of the SM, as
well as the SM contribution in the case of $C_1$. These coefficients
can be computed perturbatively. Moreover, they depend on the flavor
mixing parameters of the new physics model considered. These
parameters can therefore be constrained by experimental measurements
of observables sensitive to $\Delta S=2$ processes. The relevant
observables include the $K_L$-$K_S$ mass difference, $\Delta M_K$, and
the quantity $\epsilon$, which parameterizes indirect CP violation in
$K\to\pi\pi$ decays. The short-distance contributions to these
quantities are given by (cf. e.g. \cite{deRafael:1995zv}):
\bea
&&\Delta M_K\equiv M_{K_L}-M_{K_S}\simeq \re
\frac{\la\bar K^0|\mathcal{H}_\mathrm{eff}^\mathrm{\Delta
  S=2}|K^0\ra^*}{M_K}\labell{eq:dmk}\\[0.2cm]
&&\epsilon\equiv \frac{T[K_L\to(\pi\pi)_{I=0}]}{T[K_S\to(\pi\pi)_{I=0}]}
\simeq\frac{e^{i\pi/4}}{2\sqrt2\Delta M_K}\im
\frac{\la\bar K^0|\mathcal{H}_\mathrm{eff}^\mathrm{\Delta
  S=2}|K^0\ra^*}{M_K}\labell{eq:epsk}
\ ,
\eea
where the subscript $I$ denotes the isospin of the two pion
state. They are measured experimentally to a precision better than 1\%
(cf. \cite{Eidelman:2004wy}).

In addition to the short-distance contribution encoded in \eq{eq:dmk},
$\Delta M_K$ also receives ``long-distance'' contributions in the SM
\cite{Wolfenstein:1979wi}. These arise when the quarks which
participate in the box diagram involve the up quark. They can 
either be ``very long-distance'', involving the exchange of
pseudo-Goldstone boson pairs between two $\Delta S=1$ vertices. Or
they can be ``slightly-less long-distance'', involving the exchange of
non-Goldstone, $u,d,s$ hadrons. The former may be large, but their
determination has very large uncertainties \cite{Donoghue:1983hi}. The
latter have recently been studied in a model of large-$N_c$ QCD and
are found to be of order 10\% \cite{Cata:2004ti}. Such contributions
are not a problem in the calculation of $\epsilon$, since this
quantity is dominated by the top and charm quark contributions for
which the short distance approximation of \eq{eq:epsk} is very good. In
general, assumptions have to be made about these long distance
contributions, as well as the values of the Cabibbo-Kobayashi-Maskawa
(CKM) matrix elements in the presence of new physics. Nevertheless,
under reasonable hypotheses, interesting constraints on extensions of
the SM can be found, as shown for SUSY extensions, for instance, in
\cite{Ciuchini:1998ix}.

Regardless of the hypotheses made, to validate or constrain models of
new physics with measurements of $\Delta M_K$ and $\epsilon$, one
clearly has to be able to quantify the non-perturbative,
strong-interaction physics present in the matrix elements of the
operators $O_i$ and $\tilde O_i$ between $K^0$ and $\bar K^0$ states.
This is where a non-perturbative technique such as lattice QCD enters.
Now, while the operators in Eqs. (\ref{eq:smds2op}),
(\ref{eq:susyds2ops}) and (\ref{eq:susyds2opstilde}) have both parity
even and odd parts, it is clear that only the parity even parts
contribute to the matrix elements relevant for $K^0$-$\bar K^0$
mixing, $\la \bar K^0|O_i|K^0\ra$ and $\la \bar K^0|\tilde
O_i|K^0\ra$. Moreover, since the strong interaction preserves parity,
we have $\la \bar K^0|\tilde O_i|K^0\ra=\la \bar K^0|O_i|K^0\ra$,
$i=1,2,3$. Thus, in the remainder, we need only consider the operators
of Eqs. (\ref{eq:smds2op}) and (\ref{eq:susyds2ops}).

The transformation properties of the operators $O_i$ under the chiral
group $SU(3)_L\times SU(3)_R$ are interesting. It is commonly known
that $O_1$ belongs to the $(27,1)$ representation of this group, which
implies that its matrix element between pseudoscalar states vanishes
like the square of the mass of this state in the chiral limit, as can
be seen from its vacuum saturation value given below in
\eq{eq:b1def}. It is straightforward to work out that $O_2$ and $O_3$
belong to the $(6,6)$ representation, while $O_4$ and $O_5$ belong to the
$(8,8)$. This means that the matrix elements of these operators
between pseudoscalar states go to a constant in the chiral limit, as
indicated once again by their vacuum saturation values given below in
\eq{eq:bidef}. Thus, the matrix elements of the non-SM operators are
enhanced compared to the chirally suppressed SM one by the large
factor $M_K^2/(m_s+m_d)^2\approx 25$, as obtained for current values
of $(m_s+m_d)$ at $2\,\gev$ in the $\msbar$ scheme. Clearly, a
reliable determination of these enhanced matrix elements is needed.

The first objective of our work is to calculate the matrix elements
$\la \bar K^0|O_i|K^0\ra$, $i=1,\cdots,5$, using lattice QCD. The
calculations are performed using Neuberger fermions
\cite{Neuberger:1998bg,Neuberger:1998fp,Narayanan:1995gw,Narayanan:1994sk}
on quenched gluon backgrounds generated with the standard Wilson gauge
action. The gauge configurations were generated at two values of the
bare coupling $\beta=2N_c/g_0=6.0$ and 5.85 on $18^3\times 64$ and
$14^3\times 48$ lattices, respectively. With the lattice spacings
determined in \sec{sec:2ptfns}, i.e. $a^{-1}=2.17(9)\,\gev$ at
$\beta=6.0$ and $a^{-1}=1.49(4)\,\gev$ at $\beta=5.85$, these lattices
have sides of size 1.6 and $1.9\,\fm$. This work extends to new matrix
elements, larger lattices and to another value of the lattice
spacing--allowing for an estimate of discretization errors--our
earlier investigation of the SM matrix element $\la \bar
K^0|O_1|K^0\ra$, performed on a $16^3\times 32$ lattice at $\beta=6.0$
\cite{Garron:2003cb}. The great advantage of using Ginsparg-Wilson
fermions \cite{Ginsparg:1982bj} such as those proposed by Neuberger,
is that unlike more traditional fermion formulations (e.g. staggered
or Wilson fermions), they possess an exact, continuum-like, chiral
flavor symmetry at finite lattice spacing
\cite{Luscher:1998pq}.~\footnote{The same is true of domain-wall
fermions \protect\cite{Kaplan:1992bt} which are also Ginsparg-Wilson
fermions in the limit of an infinite fifth dimension.} In particular,
this means that the operator mixing pattern is the same as in the
continuum--a non-negligible simplification for the operators
considered here--and that discretization errors appear only at
sub-leading $a^2$ order, which at currently used lattice spacings is
certainly an advantage. Moreover, to avoid potentially large
perturbative corrections, we perform all renormalizations
non-perturbatively in the RI/MOM scheme \`a la
\cite{Martinelli:1995ty}. Preliminary results of our calculations at
$\beta=6.0$ were presented in \cite{Berruto:2003rt,Berruto:2004cm}.

Since the matrix elements are quantities of dimension four, they
suffer strongly from the uncertainty in the lattice spacing, which is
of order 10\% in quenched calculations \cite{Aoki:2002fd}. It is
therefore advantageous to give dimensionless measures of these matrix
elements, by normalizing them with quantities which are similar enough
that some of the systematic errors are expected to cancel in the
ratio. For the SM contribution, determined by the operator $O_1$, one
usually computes the $B$-parameter $B_1$ or $B_K$, which measures the
deviation of the matrix element from its vacuum saturation value. We
have
\be
B_1(\mu)=B_K(\mu)\equiv\frac{\la\bar
  K^0|O_1(\mu)|K^0\ra}{\frac{8}{3}\la\bar K^0|\bar s\gamma_\mu\gamma_5
  d|0\ra\la 0|\bar s\gamma^\mu\gamma_5 d|K^0\ra}=\frac{\la\bar
  K^0|O_1(\mu)|K^0\ra}{\frac{16}{3}M_K^2F_K^2} \ , 
\labell{eq:b1def} \ee
where $\mu$ is the renormalization scale and $M_K$ and $F_K$ are,
respectively, the mass and decay constant of the kaon, which are measured
precisely experimentally. Our normalization of the decay constant is
such that $F_K=113\,\mev$ and we take $M_K=495\,\mev$. 

One can also define $B$-parameters for the matrix elements of
$O_{2,\cdots,5}$, much in the same way. Neglecting sub-leading chiral
terms, we have:
\be
B_i(\mu)\equiv\frac{\la\bar K^0|O_i(\mu)|K^0\ra}{N_i \la\bar K^0|\bar
  s\gamma_5 d(\mu)|0\ra\la 0|\bar
  s\gamma_5 d(\mu)|K^0\ra}=-
\frac{\la\bar K^0|O_i(\mu)|K^0\ra}
{N_i\l(\frac{\sqrt2 F_KM_K^2}{m_s(\mu)+m_d(\mu)}\r)^2}
\,
\labell{eq:bidef}
\ee
for $i=2,\cdots,5$ with our phase conventions, with
$N_i=\{\frac53,-\frac13,-2,-\frac23\}$ and where we have used the
partial conservation of the axial current, $\bar s\gamma_\mu\gamma_5
d$, to rewrite the matrix elements of the pseudoscalar density $\bar
s\gamma_5 d$ in terms of the $s$ and $d$ quark masses in the second
equality. However, unlike $B_K$ which gives a dimensionless measure of
the matrix element of $O_1$ in terms of well measured quantities, the
relation of $B_{2,\cdots,5}$ to the corresponding matrix elements
involves matrix elements of $\bar s\gamma_5 d$, or equivalently
light-quark masses, which are dimensionful quantities not directly
accessible to experiment and whose determination requires a
non-perturbative calculation. Thus, in providing these $B$-parameters,
we are essentially shuffling some of the systematic errors into the
determination of the normalization factor or worse, compounding them
if the correlations between the parameters and the normalization
factors are not taken into account. This point was first emphasized in
\cite{Donini:1999nn}, where alternate suggestions for normalizing the
matrix elements were given, all involving the mass of the $K^*$.
Since the systematic errors for vector mesons may be quite different
from those on pseudoscalars, we propose another dimensionless measure
of the matrix elements, defined through the ratios:
\be
R_i^\mathrm{BSM}(\mu,M^2)
\equiv\l[\frac{F_K^2}{M_K^2}\r]_{expt}\l[\frac{M^2}{F^2}
\frac{\la\bar P^0|O_i(\mu)|P^0\ra}{\la\bar P^0|O_1(\mu)|P^0\ra}\r]_{lat},
\labell{eq:ridef}
\ee
for $i=2,\cdots,5$, where $M$ and $F$ are the mass and ``decay
constant'' of the lattice kaon which we denote by $P^0$ to indicated
that the mass of the strange and down quarks that compose it can
differ from their physical values.  The factor of $M^2$ is present to
keep the ratios from diverging in the chiral limit. Though we are not
interested in this limit here, the strong mass dependence induced by
this divergence could complicate the necessary interpolation to the
mass of the kaon. It is worth noting that the ratios
$R_i^\mathrm{BSM}(\mu,M_K^2)$ measure directly the ratio of BSM to SM
matrix elements and, as such, can be used in expressions for $\Delta
M_K$ and $\epsilon$ beyond the SM, in which the SM contribution is
factored out.

Another dimensionless measure of the matrix elements which involves only
quantities pertaining to pseudoscalar mesons are the ratios
\be
G_i(\mu,M^2)\equiv\frac{\la\bar P^0|O_i(\mu)|P^0\ra}{2F^4}
\labell{eq:gidef}
\ee
for $i=1,\cdots,5$. From these ratios, the physical matrix elements
can be obtained by considering $2F_K^4G_i(\mu,M_K^2)$. Moreover, in
the chiral limit, the $G_i$ for $i=2,\cdots,5$, once multiplied by the
correct Wilson coefficients, can be thought of as contributions to the
couplings of operators in an effective, weak chiral Lagrangian. For
simplicity, in the following we will abusively call the $G_i$
couplings.

\medskip

As already mentioned, $O_4$ and $O_5$ belong to the $(8,8)$
representation of $SU(3)_L\times SU(3)_R$. This is the same chiral
representation as the $\Delta I=3/2$ parts of the electroweak penguin
operators $Q_7$ and $Q_8$, which give the dominant $\Delta I=3/2$
contribution to $\epsilon'$. In the $SU(3)$-flavor limit, we
have:
\be
\la \pi^+|Q_{7,8}^{3/2}|K^+\ra = \frac12\la \bar K^0|O_{5,4}|K^0\ra
\ ,
\labell{eq:Q78eqO54}
\ee
where the relative phase between the two matrix
elements depends on the phase conventions chosen for pion and kaon
states~\footnote{In our conventions, the matrix elements $\la
  \pi^+|Q_{7,8}^{3/2}|K^+\ra$ differ by a sign from the ones given in
  \protect\cite{Knecht:2001bc}.} and where we define:
\bea
Q_7^{3/2}&\equiv&\frac12[\bar s^a\gamma_\mu(1-\gamma_5) d^a]
[\bar u^b\gamma^\mu(1+\gamma_5)u^b-\bar
  d^b\gamma^\mu(1+\gamma_5)d^b]\nonumber\\
&&+\frac12[\bar s^a\gamma_\mu(1-\gamma_5) u^a]
[\bar u^b\gamma^\mu(1+\gamma_5)d^b]\\
Q_8^{3/2}&\equiv&\frac12[\bar s^a\gamma_\mu(1-\gamma_5) d^b]
[\bar u^b\gamma^\mu(1+\gamma_5)u^a-\bar d^b\gamma^\mu(1+\gamma_5)d^a]
\nonumber\\
&&+\frac12[\bar s^a\gamma_\mu(1-\gamma_5) u^b]
[\bar u^b\gamma^\mu(1+\gamma_5)d^a]
\ .\eea
In this same limit we also have
\be
B^{3/2}_{7,8}=B_{5,4}
\ ,\labell{eq:B78eqB54}
\ee
where $B^{3/2}_{7,8}$ measures the deviation of $\la
\pi^+|Q_{7,8}^{3/2}|K^+\ra$ from its vacuum saturation value and here, phase
choices cancel so that the equality of \eq{eq:B78eqB54} is convention
independent.

Moreover, at leading order in the chiral expansion, these
$K^+\to\pi^+$ matrix elements are related to the
$K^0\to(\pi\pi)_{I=2}$ matrix elements relevant for the calculation of
$\epsilon'$ in the following way:
\be
\la (\pi\pi)_{I=2}|Q_{7,8}|K^0\ra\propto \frac{1}{F}
\la \pi^+|Q_{7,8}^{3/2}|K^+\ra +O(p^2)
\labell{eq:kpipiproptokpi}
\ee
where $F$ is the pseudoscalar decay constant which, at this order, can be
taken to be $F_\pi$, $F_K$ or any other value in the chiral regime and where
the constant of proportionality depends on phase and normalization
conventions. \eq{eq:kpipiproptokpi} becomes exact, of course, in the chiral
limit. 

Thus, the second objective of our work is to determine the matrix
elements $\la \pi^+|Q_{7,8}^{3/2}|K^+\ra/F$, in the chiral
limit. Ideally then, we would compute $G_{5,4}$ of \eq{eq:gidef} on
the lattice for mesons $P$ in the chiral regime and extrapolate the
result to the chiral limit to obtain the electroweak penguin couplings
$G_{7,8}^{3/2,\chi}$ of the corresponding weak, effective chiral
Lagrangian, or equivalently, the matrix element $\la
(\pi\pi)_{I=2}|Q_{7,8}|K^0\ra$ in the chiral limit. However, two
problems stand in our way. The first is that the mesons in our
simulation, except for perhaps the two to three lightest ones, are
beyond the chiral regime. The second is that the chiral behavior of
$G_{5,4}$ in the quenched and the $N_f=3$ unquenched theory differ,
indicating that we cannot expect quenched results for $G_{5,4}$ in the
chiral limit to have much to do with their physical values.

Due to computational limitations, we cannot do much about the first
problem at present, but we can try to address the second. Indeed, as
discussed in \sec{sec:massdep} and first argued in
\cite{Berruto:2004cm}, the matrix elements $\la
\pi^+|Q_{7,8}^{3/2}|K^+\ra$ can be normalized in such a way that the
resulting quantities have the same NLO chiral logarithm in the
quenched and $N_f=3$ theories. In the $SU(3)$-flavor limit, we can consider
\be
D_{7,8}^{3/2}(\mu,M^2)\equiv \frac{\la \pi^+|Q_{7,8}^{3/2}|K^+\ra}{F^2}=F^2\,G_{5,4}
\ ,\labell{eq:d78def}
\ee
where $F$ is the decay constant common to the pions and kaons. This
similarity in the chiral expansions allows for the possibility that
the quenched and unquenched results for these quantities in the chiral
limit may not be too different.

\medskip

Because of its importance in constraining the CP violating phase of
the Cabibbo-Kobayashi-Maskawa (CKM) matrix with the measurement of
$\epsilon$ (cf. e.g. \cite{Charles:2004jd,Bona:2005vz}), $B_K$ has
received much attention, both analytically and in lattice
QCD. Analytically, it has been evaluated with two-point
\cite{Pich:1985ab,Prades:1991sa} and three-point
\cite{Chetyrkin:1985vj,Decker:1986tx,Bilic:1987gk} QCD sum-rules, as
well as in models of large $N_c$ QCD
\cite{Bijnens:1995br,Peris:2000sw,Bijnens:2006mr}. The most recent
calculations were performed in the chiral limit, except for that of
\cite{Bijnens:1995br}, where it was shown that chiral corrections
represent approximatively 50\% of the final number. On the lattice
$B_K$ also has a long history, which has been reviewed recently in
\cite{Lellouch:2002nj,Hashimoto:2004hn,Dawson:2005za}. In particular,
there are at present two quenched benchmark calculations
\cite{Aoki:1998nr,Dimopoulos:2006dm} and, very recently, the first
$N_f=2+1$ unquenched calculation with staggered fermions, at a single
value of the lattice spacing \cite{Gamiz:2006sq}. The point of our
calculation is not to establish a new quenched benchmark for $B_K$,
but rather to quantify $K^0$-$\bar K^0$ mixing beyond the SM in a
consistent framework. In fact, our dimensionless ratios
$R_i^\mathrm{BSM}$ of \eq{eq:ridef}, in which a number of systematic
errors are expected to cancel, may be used in the future, in
conjunction with high precision unquenched determinations of $B_K$, to
yield improved calculations of $K^0$-$\bar K^0$ mixing beyond the SM.

Unlike the situation that we have described in the preceding
paragraph for $B_K$, there is only one other lattice calculation of
the full set of $\Delta S=2$ matrix elements relevant for $K^0$-$\bar
K^0$ mixing~\cite{Donini:1999nn}. It was performed with
tree-level $O(a)$-improved Wilson fermions, which are not chirally
symmetric and suffer for potentially large $O(\alpha_sa)$ errors. In fact
we find results for the BSM matrix elements which are
substantially larger than those of \cite{Donini:1999nn}, implying
stronger constraints on BSM models than previously found.

The situation regarding electroweak penguin matrix elements is
somewhat intermediate. There are a number of analytical calculations
\cite{Narison:2000ys,Knecht:2001bc,Bijnens:2001ps,Cirigliano:2002jy}
as well as a handful of recent quenched lattice calculations
\cite{Pekurovsky:1998jd,Noaki:2001un,Blum:2001xb,DeGrand:2003in,
Boucaud:2004aa} and a preliminary $N_f=2$ unquenched calculation
\cite{Noaki:2005zw}. There are also older calculations of the
corresponding $B$-parameters \cite{Gupta:1996yt,Kilcup:1997ye} as well
as exploratory calculations which attempt to go beyond the chiral
limit \cite{Boucaud:2004aa,Noaki:2005zw}, including one
\cite{Yamazaki:2005eg} which is based on the finite-volume formalism
of \cite{Lellouch:2000pv}. Yet there is no consensus on the value of
these matrix elements even in the chiral limit. Thus, our results,
obtained in a formulation of lattice QCD with full chiral symmetry,
full $O(a)$-improvement, at two values of the lattice spacing and
using non-perturbative renormalization, are useful for clarifying the
current situation.

Our paper is organized as follows. In \sec{sec:numerics} we briefly
summarize the parameters of our simulation. In \sec{sec:2ptfns} we
describe our analysis of two-point functions relevant for the
determination of the matrix elements of interest. This includes a
description of our extraction of the strange quark mass. In
\sec{sec:3ptfns} we use three-point functions to determine the bare
$B$-parameters and other quantities related to the four-quark operator
matrix elements, as functions of our bare quark masses. \sec{sec:npr}
is dedicated to the non-perturbative calculation of the
renormalization constants required to subtract the logarithmic
divergences present in the bare results. In \sec{sec:massdep} we
discuss the quark-mass dependence of the results renormalized
conventionally at $2\,\gev$~\footnote{The renormalization scale
$\mu=2\,\gev$ is convenient, because it is in the perturbative domain
and because $2\,\gev$ is comparable to the lattice spacings that we
use. The latter guarantees that we do not introduce potentially large
$N_f$-dependent $\ln(a\mu)$ logarithms in matching our results to
continuum renormalization schemes.} in the RI/MOM scheme and discuss
the necessary interpolations and extrapolations to the physical quark
masses. As in the previous sections, the results obtained at the two
values of the lattice spacing are discussed in parallel, and a direct
comparison of these results is presented. In \sec{sec:finalres} we
combine the results obtained at the two values of the lattice spacing
for the matrix elements, matrix element ratios and quark masses
interpolated or extrapolated to their physical mass point. We give our
final physical results at $2\,\gev$ in both the RI/MOM and
$\msbar$-NDR schemes, including estimates of the dominant systematic
errors within the quenched approximation. We also compare our results
to those obtained both on and off the lattice. The reader interested
only in our final results can skip directly to this section. Finally,
in \sec{sec:ccl}, we conclude with a few closing remarks.

The paper further contains a number of
appendices. \apps{sec:app2ptfns}{sec:appmassdep} contain tables which
are referred to in earlier sections, but which we place in appendices
to improve the readability of the paper. For completeness, we
provide in \app{sec:appndrrenormcsts} the non-perturbative
renormalization constants in the $\msbar$-NDR scheme as well as the
quark-mass dependence of the four-quark matrix element results
renormalized in the $\msbar$-NDR scheme at $2\,\gev$ in
\app{sec:appmassdepndr}. We also give, in \app{sec:appndrres}, results
in the $\msbar$-NDR scheme at the individual lattice spacings for the
$B$-parameters and other matrix element ratios at the physical
point. These results are used in \sec{sec:finalres} to obtain our
final physical results in that scheme.

\section{Numerical implementation}

\labell{sec:numerics}

We describe gluons with the Wilson gauge action and quarks with
the overlap action
\be
S=a^4\sum_x\bar q D[m_q] q=a^4\sum_x\bar q\l[\l(1-\frac1{2\rho}am_q\r)
D[0]+m_q\r]q
\ ,\ee
where $m_q$ is the bare quark mass and $D[0]$ is the Neuberger-Dirac
operator~\cite{Neuberger:1998fp}
\be
D[0]=\frac{\rho}{a}\l(1+X/\sqrt{X^\dagger X}\r)
\ .
\labell{eq:DN0}\ee
In \eq{eq:DN0}, $X=D_W-\rho/a$ is the Wilson-Dirac operator with a
negative mass parameter $\rho$ which separates physical from doubler
modes. We generated 100 quenched gauge configurations each at a bare
value of the gauge coupling, $\beta=6.0$ and
$\beta=5.85$, corresponding to inverse lattice spacings of
$a^{-1}=2.12\,\gev$, respectively $a^{-1}=1.61\,\gev$, as obtained from the
Sommer scale defined by $r_0^2F(r_0)=1.65$ and $r_0=0.5\,\fm$
\cite{Sommer:1994ce,Guagnelli:1998ud}.~\footnote{As detailed in
\protect\sec{sec:2ptfns}, compatible values for the lattice spacing
are obtained from more physical quantities, such as the leptonic decay
constant of the kaon, $F_K$.} On each gauge configuration, we invert
the Neuberger-Dirac operator for a number of values of the quark mass
ranging from roughly $m_s/3$ to $1.5 m_s$. At $\beta=6.0$, the bare
masses in lattice units are $am_q=0.030, 0.040, 0.060, 0.080, 0.100$
and at $\beta=5.85$, they are $am_q=0.030, 0.040, 0.053, 0.080, 0.106,
0.132$.  Details on our implementation of the Neuberger-Dirac operator
can be found in \cite{Babich:2005ay}. All statistical errors are
estimated with a single-elimination jackknife, which is propagated
throughout our analysis.

In our simulations, we choose the negative mass parameter $\rho$ to
vary with the lattice spacing $a$, namely $\rho=1.4$ at $\beta=6.0$
and $\rho=1.6$ at $\beta=5.85$, to optimize the locality of the
Neuberger-Dirac operator~\cite{Hernandez:1998et}. At fixed lattice
spacing, a change in $\rho$ corresponds to an $O(a^2)$ redefinition of
the action and one may worry about the effect on the continuum limit
of our procedure.  However, as long as $\rho$, viewed as a function
of the lattice spacing $a$, remains in the single particle sector and
has a smooth limit in the range $0<\rho<2$ for $a\to 0$, changing
$\rho$ with $\beta$ simply represents a modification of the approach
to the continuum limit. Moreover, for a function $\rho$ which
approaches its continuum limit value with corrections of $O(a^2)$,
this modification only appears at $O(a^4)$. We require that our two
values of $\rho$ lie on such a curve. 

\section{Two-point functions, spectral quantities and scale setting}

\labell{sec:2ptfns}

To obtain the vacuum saturation values of the matrix elements of interest, we
compute the following two-point functions:
\be
C^{(2)}_{PP}(x_0)=\sum_{\vec{x}}\la P(x)\bar P(0)\ra
\labell{eq:cppdef}
\ee
and
\be
C^{(2)}_{PA_0}(x_0)=\sum_{\vec{x}}\la P(x) \bar A_0(0)\ra
\labell{eq:cpaodef}\ ,\ee
where $P\equiv\bar d\gamma_5\hat s$ is the pseudoscalar density and
$A_0$ is the time-component of the axial current $A_\mu\equiv\bar
d\gamma_\mu\gamma_5\hat s$. Here $\hat s=(1-a D[0]/2\rho)s$. The
introduction of the ``hatted'' quark fields guarantees that the
bilinear and four-quark operator matrix elements have the proper
chiral properties and therefore renormalize as in the continuum and
are free of $O(a)$ discretization errors. Moreover $\bar P$ and $\bar
A_0$ are obtained from $P$ and $A_0$ by interchanging the flavors
$d\leftrightarrow s$. We work in the $SU(3)$-flavor limit where
$m_s=m_d=m_q$ and we allow the common bare quark mass $am_q$ to take
on the values described in the preceding section.

While the correlation function $C^{(2)}_{PA_0}(x_0)$ is only used to
normalize the three-point function involving operator $O_1$ as
described below, we fit the time-symmetrized correlation function
$C^{(2)}_{PP}(x_0)$ to the usual cosh form in the time interval $12\le
x_0/a\le 19$ at $\beta=6.0$ and $10\le x_0/a\le 14$ at $\beta=5.85$
for all quark mass values, obtaining the masses $M$ of the
corresponding pseudoscalar mesons as well as the matrix elements $\la
0|P|P^0\ra$. The initial fit time ensures that asymptotic behavior has
been reached while the final time is chosen to match the range used in
three-point function fits, as described in \sec{sec:3ptfns}. Because
of the chiral symmetry of Neuberger fermions, the axial Ward identity
can be used to derive the pseudoscalar decay constant $F$ from this
matrix element, without a finite renormalization.  Results for $aM$
and $aF$ are summarized in \tab{tab:600spectlatunits} for $\beta=6.0$
and \tab{tab:585spectlatunits} for $\beta=5.85$.

Because Neuberger fermions satisfy an index theorem
\cite{Hasenfratz:1998ri,Luscher:1998pq}, one may worry about the
presence of finite-volume, quenched, zero mode contamination in the
$C^{(2)}_{PP}(x_0)$ correlator. However, we showed in
\cite{Babich:2005ay} that these contributions are not visible within
our statistical accuracy at the volumes and quark masses that we
consider.  In \cite{Babich:2005ay}, we compared directly results of
fits to $C^{(2)}_{PP}(x_0)$ and $C^{(2)}_{(P+S)(P-S)}(x_0)$, where $S$
is the scalar density, and found no appreciable difference. The two
correlators are both dominated by the pseudo-Goldstone boson state for
asymptotic times $x_0$ and $C^{(2)}_{(P+S)(P-S)}(x_0)$ is free from
zero-mode contributions by chirality.  For completeness, this
comparison is repeated here. Results obtained for the time-symmetrized
correlation function $C^{(2)}_{(P+S)(P-S)}(x_0)$, for the same time
fitting ranges as above, are also given in
\tabs{tab:600spectlatunits}{tab:585spectlatunits}.

As \tabs{tab:600spectlatunits}{tab:585spectlatunits} show,
$C^{(2)}_{PP}(x_0)$ yields results for $aM$ and $aF$ which are
compatible with those given by $C^{(2)}_{(P+S)(P-S)}(x_0)$. The
compatibility is also displayed graphically in
\figs{fig:latplane600}{fig:latplane585}, where the results for $aF$
from the two correlation functions are plotted against the
corresponding values $M^2/(4\pi F)^2$. We conclude that the effects of
finite-volume zero-mode contributions are not statistically significant
at the volumes and quark masses that we consider, at least for
two-point functions at asymptotic times. For simplicity, we will
consider only $C^{(2)}_{PP}(x_0)$ in the remainder of this paper.

\begin{figure}[t]
\begin{center}
\psfig{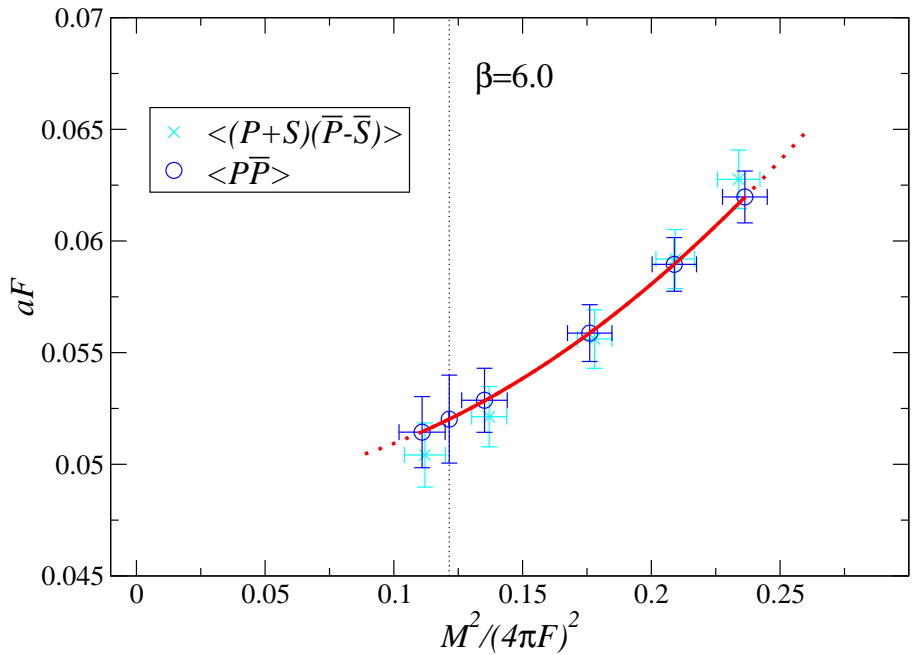}
\caption{\labell{fig:latplane600} Mass-dependence at $\beta=6.0$, in
  terms of the variable $M^2/(4\pi F)^2$, of the results for the decay
  constant in lattice units, $aF$, obtained from the two correlation
  functions $C^{(2)}_{PP}(x_0)$ and $C^{(2)}_{(P+S)(P-S)}(x_0)$. The
  solid curve is a quadratic fit to the results from
  $C^{(2)}_{PP}(x_0)$. It is used to interpolate the results to the
  kaon point $M^2/(4\pi F)^2=M_K^2/(4\pi F_K)^2$, shown as a vertical
  dotted line.  The result of this interpolation is then used to
  determine the lattice spacing.}
\end{center}
\end{figure}

\begin{figure}[t]
\begin{center}
\psfig{file=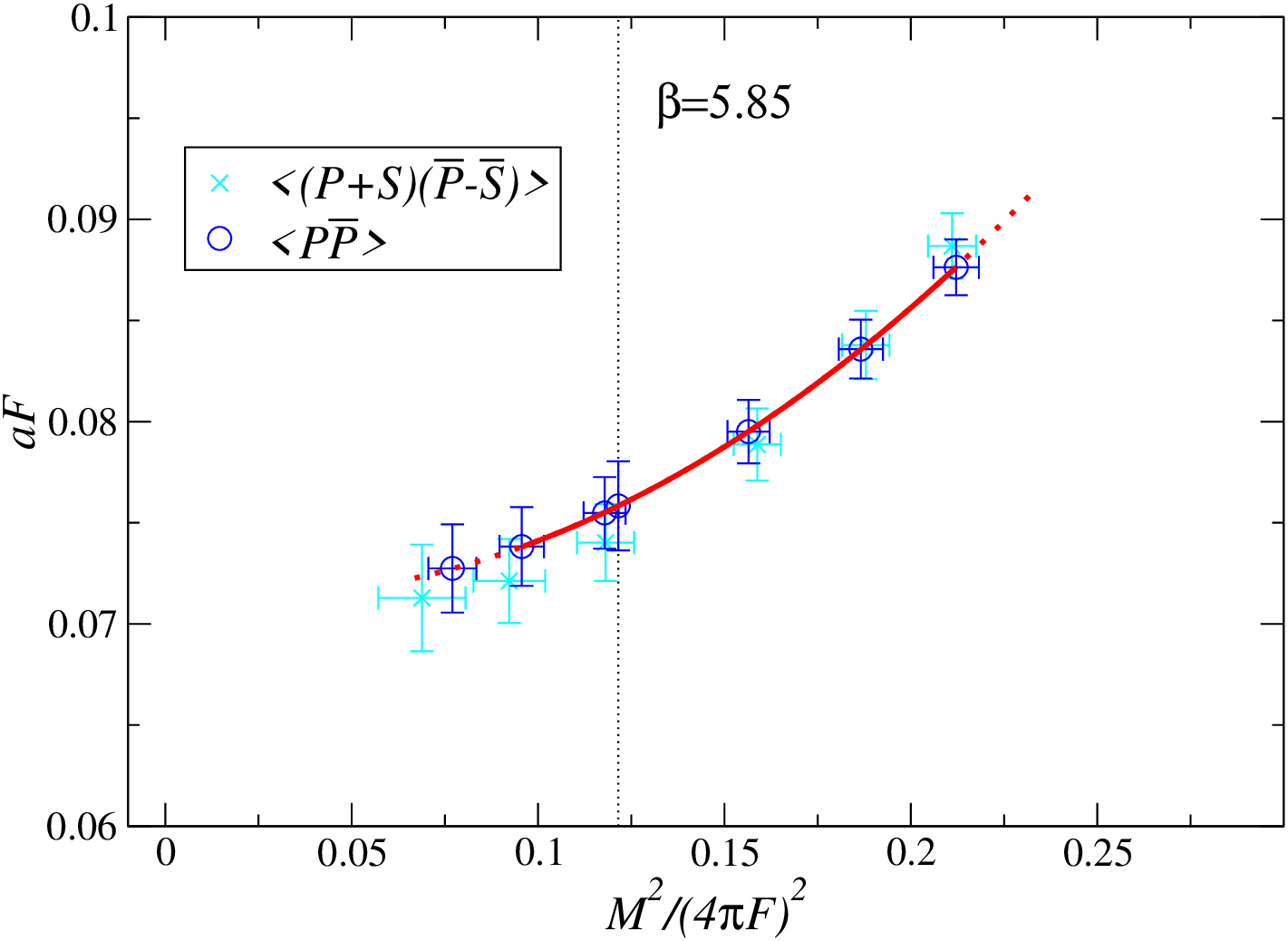,width=12cm}
\caption{\labell{fig:latplane585} Same as \fig{fig:latplane600}, 
but for $\beta=5.85$.}
\end{center}
\end{figure}

To determine the lattice spacing $a$, we study $aF$ as a function of
$M^2/(4\pi F)^2$ and interpolate $aF$ quadratically to the point
$M^2/(4\pi F)^2=M_K^2/(4\pi F_K)^2$, with $M_K$ and $F_K$ given in the
Introduction. The interpolations are shown in \fig{fig:latplane600}
for $\beta=6.0$ and in \fig{fig:latplane585} for $\beta=5.85$, where
the result corresponding to $am_{q}=0.030$ is omitted because
$M_PL\simeq 3.5$ for this point, which makes it more susceptible to
finite volume effects. With the value of the decay constant in lattice
units obtained at the kaon point, we set $a^{-1}=F_K/(aF)$ and obtain
$a^{-1}=2.17(9)\,\gev$ at $\beta=6.0$ and $a^{-1}=1.49(4)\,\gev$ at
$\beta=5.85$.  These values for the lattice spacing are compatible
with those derived from the Sommer scale in \sec{sec:numerics}, namely
$a^{-1}=2.12\,\gev$ at $\beta=6.0$ and $a^{-1}=1.61\,\gev$ at
$\beta=5.85$, given the intrinsic $O(10\%)$ model-dependence
associated with $r_0$.  We choose to set the scale and fix the strange
quark mass with $M_K$ and $F_K$ because these quantities are closely
related to the matrix elements of interest. This procedure may
therefore absorb some of the quenching artefacts associated with the
physics that we are studying. In any event, we will be considering
mainly dimensionless quantities such as $B$-parameters and
matrix-element ratios. The value of the lattice spacing will then only
enter logarithmically through the renormalization scale dependence of
the matrix elements and in fixing the kaon point in the mild
interpolations that we will have to perform.

As a side product of our analysis, we determine the value of the
strange quark mass. The renormalized quark mass is obtained from the
bare quark mass in lattice units as
$m^\RI(2\,\gev)=(am)*a^{-1}/Z_S^\RI(2\,\gev)$, where the lattice
spacing is the value determined above, and $Z_S^\RI(2\,\gev)$ is the
renormalization constant of the scalar density determined
non-perturbatively in \sec{sec:npr}. We assume that the meson mass
depends only on the sum of the masses of the quarks which compose
it. In \fig{fig:msplusmhatRI600and585}, we show a quadratic
interpolation to the kaon point of the RI/MOM sum of quark masses at
$2\,\gev$, $(m_{q_1}+m_{q_2})^\RI(2\,\gev)$, as a function of
$M^2/(4\pi F)^2$ at $\beta=6.0$. Also shown in this plot are the
results at $\beta=5.85$, where the interpolation, performed without
the point corresponding to $am_{q}=0.030$ for the reasons discussed
above, has been omitted for clarity. Only statistical errors are
shown, and already the agreement is very good.  The interpolation to
the kaon point gives $(m_s+\hat m)^\RI(2\,\gev)$, where $\hat m$ is
the average up and down quark mass. Our results for this quantity,
obtained at the two values of the lattice spacing are summarized in
\tab{tab:msplusmhatRI600and585}. There is no evidence for
scaling violations. Results in the $\msbar$ scheme are given in
\app{sec:appndrres}. The final result will be given in
\sec{sec:finalres}.

\begin{table}[t]
\begin{center}
\caption{\labell{tab:msplusmhatRI600and585}
$(m_s+\hat m)$ at $2\,\gev$ in the $\RI$ scheme as
  obtained from our simulations at $\beta=6.0$ and $\beta=5.85$.
The first error is
statistical and the second comes from the systematic uncertainty in the
determination of $Z_S$. }
\begin{tabular}{cc}
\hline
\hline
$\beta$ & $(m_s+\hat m)^\RI(2\,\gev)$\\
\hline
6.0 & $120.(8)(2)\,\mev$\\
5.85 & $127.(6)(10)\,\mev$\\
\hline
\hline
\end{tabular}
\end{center}
\end{table}

\begin{figure}[t]
\begin{center}
\psfig{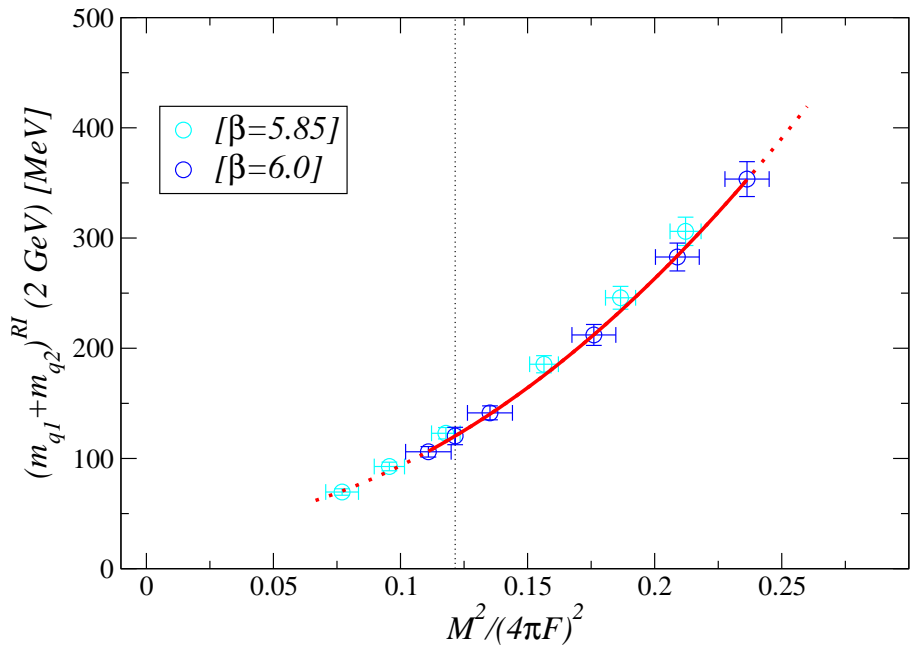}
\caption{\labell{fig:msplusmhatRI600and585} Mass-dependence at
  $\beta=6.0$, in terms of the variable $M^2/(4\pi F)^2$, of the sum
  of the quark masses which compose the pseudoscalar meson of mass
  $M$. The quark masses are renormalized in the RI/MOM scheme at
  $2\,\gev$. The solid curve is the result of a quadratic fit, plotted
  in the fit region. The fit is used to interpolate the results to the
  kaon point $M^2/(4\pi F)^2=M_K^2/(4\pi F_K)^2$, shown as a vertical
  dotted line.  The dashed curve is an extension of the fit curve
  outside the fit range. Also shown are the results at $\beta=5.85$,
  but the interpolation is omitted for clarity. Only statistical
  errors are shown.}
\end{center}
\end{figure}

\section{Three-point functions, bare $B$-parameters and four-quark operator
  matrix elements}

\labell{sec:3ptfns}

To obtain the bare matrix elements of the $\Delta S=2$ operators,
$O_i$, between $K^0$ and $\bar K^0$ states, and the corresponding
$B$-parameters, we construct the following ratios of three- to
two-point functions:
\bea
\cB^1_{PP}(x_0,y_0)&=&
\frac{\sum_{\vec{x},\vec{y}}{\la P(x) O_1(0) P(y)\ra}}
{\frac83
\sum_{\vec{x},\vec{y}}{\la P(x)\bar A_0(0)\ra\la \bar A_0(0)P(y)\ra}}
\stackrel{a\ll x_0\ll T/2\ll y_0\ll T}{\longrightarrow}
B_1\labell{eq:3pt12pt2pt}\\
\cB^i_{PP}(x_0,y_0)&=&\frac{\sum_{\vec{x},\vec{y}}{\la P(x) O_i(0) P(y)\ra}}
{N_i\sum_{\vec{x},\vec{y}}{\la P(x)\bar P(0)\ra\la \bar P(0)P(y)\ra}}
\stackrel{a\ll x_0\ll T/2\ll y_0\ll T}{\longrightarrow}
B_i
\labell{eq:3pti2pt2pt}
\ ,
\eea
for $i=2,\cdots,5$, with the $N_i$'s given after \eq{eq:bidef}. As
indicated, these ratios become constant for asymptotically large
initial ($y_0$) and final ($x_0$) times. These constants are the bare
$B$-parameters, $B_i$, $i=1,\cdots,5$, whose renormalized analogs
were defined in \eq{eq:bidef}. In \eqs{eq:3pt12pt2pt}{eq:3pti2pt2pt},
the subscripts on $\cB^i_{J_fJ_i}$, $i=1,\cdots,5$, label the sink
$J_f(x)$ and the source $J_i(y)$, which we defined below
\eq{eq:cpaodef} for the case of $P(x)$ and $A_0(x)$. In addition, the
bare operators $O_i$ in \eqs{eq:3pt12pt2pt}{eq:3pti2pt2pt} are those
of \eqs{eq:smds2op}{eq:susyds2ops} with $d$ replaced by $\hat d$. 

We fit the $\cB^i_{PP}$ ratios of \eqs{eq:3pt12pt2pt}{eq:3pti2pt2pt}
to a constant in the symmetric time intervals given by: $12\le
x_0/a\le 19$ and $45\le y_0/a\le 52$ for $i{=}1,\cdots,5$, at
$\beta=6.0$; $10\le x_0/a\le 12$ and $36\le y_0/a\le 38$ for $i{=}1$
and $10\le x_0/a\le 14$ and $34\le y_0/a\le 38$ for $i{=}2,\cdots,5$,
at $\beta=5.85$. The points nearest the origin are are chosen so that
the ratios are dominated by the ground state contribution. The other
extremities of the ranges are picked so that time reversed
contributions are less than 1\%, even for the lightest meson (next to
lightest at $\beta=5.85$). The results for the bare $B$-parameters are
given in \tab{tab:bareB600} at $\beta=6.0$ and \tab{tab:bareB585} at
$\beta=5.85$, as functions of bare quark mass.

Though no evidence was found in the previous section for
finite-volume, zero-mode contamination in two-point functions, one
could worry about their effect on three-point functions. In order to
verify whether these modes might be a problem, we considered an
alternate ratio for determining $B_1$ which, unlike
$\cB^1_{PP}(x_0,y_0)$ of \eq{eq:3pt12pt2pt}, is entirely free of
zero-mode contamination because of its chiral structure. This ratio is
the one that we advocated in our first determination of $B_K$ with
Neuberger fermions \cite{Garron:2003cb}, performed on a lattice whose
four-volume is roughly three times smaller than the one considered
here:
\be
\cB^1_{L_0L_0}(x_0,y_0)=
\frac{\sum_{\vec{x},\vec{y}}{\la L_0(x) O_1(0) L_0(y)\ra}}
{\frac83
\sum_{\vec{x},\vec{y}}{\la L_0(x)\bar L_0(0)\ra\la \bar L_0(0)L_0(y)\ra}}
\stackrel{a\ll x_0\ll T/2\ll y_0\ll T}{\longrightarrow}
B_1\labell{eq:lolo3pt12pt2pt}
\ ,\ee
where $L_0$ is the time-component of the left-handed weak current,
$L_\mu=\bar d\gamma_\mu(1-\gamma_5)\hat s$.  As found in
\cite{Garron:2003cb}, this ratio reaches asymptotic behavior at early
times. Fits of this ratio in symmetric time ranges with $x_0/a\ge 8$
and $y_0/a\le 56$ at $\beta=6.0$ and with $x_0/a\ge 6$ and $y_0/a\le
42$ at $\beta=5.85$ gives bare values of $B_1$ which are entirely
compatible with those obtained from $\cB^1_{PP}(x_0,y_0)$.  We take
this as evidence that the three-point functions which appear in
\eqs{eq:3pt12pt2pt}{eq:3pti2pt2pt} are not significantly affected by
zero-mode contributions for the masses that we consider in our
relatively large volumes.

In computing matrix elements of the operators $O_2$ and $O_3$, it is
also possible to find sources which entirely eliminate finite-volume,
zero-mode contributions.  In the definition of
$\cB^{2,3}_{PP}(x_0,y_0)$ in \eq{eq:3pti2pt2pt}, one must replace the
pseudoscalar source and sink in the numerator by $R\equiv S+P$, where
$S$ is the scalar density, $S\equiv\bar d\hat s$. This eliminates
zero-mode contributions in the three-point function by chirality. In
the denominator, the two point function $\la P(x)\bar P(0)\ra$ must be
replaced by $\la R(x)\bar L(0)\ra$, and $\la \bar P(0)P(y)\ra$ by
$\la\bar L(0)R(y)\ra$, where $L\equiv S-P$. Both these two-point
functions are free from zero modes. The resulting ratios,
$\cB^{2,3}_{RR}(x_0,y_0)$ are thus also free from zero-mode
contributions. As one can easily convince oneself, such a manipulation
is not possible in correlation functions relevant for computing the
matrix elements of the operators $O_4$ and $O_5$. There, zero-mode
contributions cannot be eliminated by simply using
chirality. Nevertheless, in \cite{Berruto:2004cm}, we have checked
that sources with different chiral properties, and thus different
zero-mode contributions, give identical results. In any case, we do
not pursue this issue further here and assume that the absence of
zero-mode contributions observed in pseudoscalar two-point functions
and in the three-point function of \eq{eq:3pt12pt2pt} is evidence that
these contributions are not a problem. And we take as our final
results for the $B$-parameters those given by the ratios of
\eqs{eq:3pt12pt2pt}{eq:3pti2pt2pt}, because the use of identical
sources for the different operator matrix elements yields
cancellations of some statistical fluctuations in the ratio of these
matrix elements.

\bigskip

From these $B$-parameters and our determinations of $M$ and $F$ from the
two-point function $C^{(2)}_{PP}$, we then derive a number of other quantities
of interest. The building blocks for these quantities are the bare matrix
elements of our five operators $O_i$, $i=1,\cdots,5$, which we obtain as
follows:
\bea
\la \bar P^0|O_1|P^0\ra&=&\frac{16}{3Z_A^2}M^2F^2 B_1\labell{eq:baremeo1}\\
\la \bar P^0|O_i|P^0\ra&=&-N_i|\la 0|P|P^0\ra|^2 B_i
\ .\labell{eq:baremeoi}
\eea
In \eq{eq:baremeo1}, $Z_A$ is the normalization constant of the local axial
current, $A_\mu$, defined after \eq{eq:cpaodef}. It is there because the vacuum
saturation value of the matrix element of $O_1$, used in the definition of
$B_1$, is obtained in terms of the matrix element of $A_0$, while in
constructing the matrix element in \eq{eq:baremeo1}, we have used the
partially conserved axial current, through the axial Ward identity. $Z_A^2$
could be omitted here and reintroduced when we renormalize our results.
However, for our simulation parameters it leads to a large correction
(see \tab{tab:zares}). We have thus chosen to introduce
it, because it allows for direct comparison of quantities containing $\la
\bar P^0|O_1|P^0\ra$ of \eq{eq:baremeo1} to those that are obtained using a
vacuum saturation calculated with matrix elements of the local axial current
$A_0$.

In fact, we have performed such a comparison using $\la 0|A_0|P^0\ra$
obtained from both $C^{(2)}_{PA_0}$ and $C^{(2)}_{A_0P}$, together
with $C^{(2)}_{PP}$.  The bare matrix elements determined using the
three different estimates of its vacuum saturation value are entirely
compatible. However, the matrix element's statistical error does
depend quite strongly on the procedure used to obtain it from
$B_1$. The matrix element obtained using a vacuum saturation value
from $C^{(2)}_{PA_0}$ is the one with the smallest statistical
error. It is followed by the one obtained which makes use of
$C^{(2)}_{A_0P}$ and finally by the one of \eq{eq:baremeo1}, which
makes use of only $C^{(2)}_{PP}$.  Nonetheless, the best measure of
the physical matrix element $\la \bar K^0|O_1|K^0\ra$ is given by
$\frac{16}{3}M_K^2F_K^2B_1$, which does not necessitate
reconstructing $\la \bar P^0|O_1|P^0\ra$ for our lattice mesons,
$P^0$.  Moreover, obtaining $\la \bar P^0|O_1|P^0\ra$ from $B_1$ and a
fit to $C^{(2)}_{PA_0}$ gives results for the BSM ratios of
\eq{eq:ridef} which have significantly larger statistical errors than
those obtained using \eq{eq:baremeo1}. The same is true of the matrix
element obtained from $B_1$ and a fit to $C^{(2)}_{A_0P}$. In fact,
both these determinations of BSM ratios have very comparable
errors. We thus take $\la \bar P^0|O_1|P^0\ra$ as obtained through
\eq{eq:baremeo1} and avoid discussing fits of the two-point functions
$C^{(2)}_{PA_0}$ and $C^{(2)}_{A_0P}$ which have already been detailed
in \cite{Babich:2005ay}. 

In addition, as discussed in the Introduction, we have chosen not to
give the values of the matrix elements themselves, which have mass
dimension four and are therefore very sensitive to the intrinsic
uncertainty associated with the determination of the lattice spacing
in quenched calculations. Rather we choose to present two
dimensionless measures of these matrix elements, namely the BSM ratios
of \eq{eq:ridef} and the couplings of \eq{eq:gidef}, the latter
enabling a determination of the matrix element itself for a given
meson mass, using the corresponding value of the leptonic decay
constant.

Thus, from the matrix elements obtained using
\eqs{eq:baremeo1}{eq:baremeoi} and the values of $M$ and $F$ from fits
to $C^{(2)}_{PP}$ (\tabs{tab:600spectlatunits}{tab:585spectlatunits}),
we straightforwardly construct the bare equivalents of the BSM ratios
defined in \eq{eq:ridef}.  We summarize our results for these ratios
as a function of bare quark mass at $\beta=6.0$ in
\tab{tab:ribarevsam600} and at $\beta=5.85$ in
\tab{tab:ribarevsam585}.

We also construct the bare equivalents of the couplings $G_i$ of
\eq{eq:gidef}, from the matrix elements as given by
\eqs{eq:baremeo1}{eq:baremeoi}, and the pseudoscalar decay constant
obtained from $C^{(2)}_{PP}$ and the axial Ward identity and given,
again, in \tabs{tab:600spectlatunits}{tab:585spectlatunits}. Our results
for these contributions to the bare couplings as a function of bare
quark mass are summarized in \tab{tab:gibarevsam600} for $\beta=6.0$
and \tab{tab:gibarevsam585} for $\beta=5.85$.

Finally, in view of obtaining the chiral limit values of the matrix
element of the electroweak penguin operators relevant for direct CP
violation in $K\to\pi\pi$ decays, we determine the quantities
$D_{7,8}^{3/2}$, defined in \eq{eq:d78def}. Our results for these
quantities in lattice units are summarized in
\tab{tab:a2dibarevsam600} for $\beta=6.0$ and
\tab{tab:a2dibarevsam585} for $\beta=5.85$.

\section{Non-perturbative renormalization}

\labell{sec:npr}

The bare $\Delta S=2$ matrix elements of operators $O_i$, $i=1,\ldots,
5$, have logarithmic divergences which must be subtracted. Because
Neuberger fermions exhibit, at finite lattice spacing, a chiral flavor
symmetry analogous to that of continuum QCD, the renormalization goes
through as in the continuum. In particular, $O_1$, the standard model
operator renormalizes multiplicatively, while $O_2$ and $O_3$ mix
under renormalization, as do $O_4$ and $O_5$. These mixing properties
are also compatible with the chiral properties of the operators
described in the Introduction.

To avoid potentially large perturbative errors, we compute all
renormalization constants non-perturbatively in the RI/MOM scheme \`a
la \cite{Martinelli:1995ty}. Thus, we fix gluon configurations to
Landau gauge and numerically compute the relevant amputated forward
quark Green functions with legs of four-momentum $p$,
$\Lambda_{O_i}(m_q,p,g_0)$, where $m_q$ is the mass of the quark.  The
renormalization constants are then determined by requiring that the
corresponding renormalized vertex functions have their tree-level
values.  We therefore define the ratios:
\begin{equation}
\cR^{\RI}_{ij}(m_q,p,g_0)\equiv Z_A^2\frac{\Tr\l\{\Lambda_V(m_q,p,g_0)
\cP_{V}\r\}^2}{\Tr\l\{\Lambda_{O_j}(m_q,p,g_0) \cP_{O_i}\r\}}\ ,
\labell{eq:rirats}
\end{equation}
where $i=j=1$ for $O_1$, $i,j\in\{2,3\}$ to determine the mixing of
$O_2$ and $O_3$ and $i,j\in\{4,5\}$ for the mixing pair
${O_4,O_5}$. In \eq{eq:rirats}, the $\cP_\cO$ are normalized
projectors onto the spin-color structure of the tree-level operator
$\cO$ \cite{Donini:1999sf}. Because of the chiral symmetry of
Neuberger fermions, the parity even and odd parts of the operators
renormalize in the same way. We therefore consider the parity odd
parts, as CPS symmetry then guarantees that off-diagonal mixing terms
vanish exactly.

Chiral symmetry also guarantees that the product $Z_A\times
\Tr\l\{\Lambda_V(m_q,p) \cP_{V}\r\}$ is the quark wavefunction
renormalization squared, $Z_q$. Since we can obtain $Z_A$ very cleanly
from the axial Ward identity \cite{Babich:2005ay}, this method allows
us to circumvent a more direct calculation of $Z_q$, a procedure
which induces large discretization errors.  We choose $\Lambda_V$
instead of $\Lambda_A$ because the former has a slightly cleaner
signal. In \tab{tab:zares}, we summarize the results for $Z_A$
obtained from the axial Ward identity.

\begin{table}[t]
\caption{\labell{tab:zares} Results for the renormalization constant
$Z_A$ of
the local axial current at $\beta=6.0$ and $\beta=5.85$.}
\begin{tabular}{cc}
\hline
\hline
$\beta$ & $Z_A$\\
\hline
6.0 & 1.5548(11)\\
5.85 & 1.4434(18)\\
\hline
\end{tabular}
\end{table}

The momentum combinations that we consider in constructing the ratios
of \eq{eq:rirats} are the following. We allow
$p_{1,2,3}\times(L/2\pi)$ to take on values from 0 to 2, and
$p_0\times(T/2\pi)$, values from 0 to 4. For larger momenta, we
restrict $p_{1,2,3}\times(L/2\pi)$ to take on values from 2 to 3 and
$p_0\times(T/2\pi)$, values from 7 to 10.  Moreover, we use the
lattice-fermion definition of momentum, $p^2\equiv
(1/a^2)\sum_{\mu=0}^3 (\sin ap_\mu)^2$.  These choices reduce
$O(4)$-breaking discretization errors so that the
$\cR^{\RI}_{ij}(m_q,p,g_0)$ are to a good approximation functions of
$p^2$.

At large $p^2$ ($p^2 \gsim 1.5\,\gev^2$), an operator product expansion (OPE)
in $1/p^2$ can be performed on the Green functions which appear in the
definition of $\cR^{\RI}_{ij}(m_q,p,g_0)$. In this expansion, the leading
term, proportional to the unit operator, gives the desired renormalization
constants in the RI/MOM scheme. Perturbative mass effects appear at sub-leading
orders, suppressed by powers of $p^2$, as do possible non-perturbative
contributions. On the lattice, of course, $O(4)$ symmetry is broken and there
will be discretization errors. However, as already noted, for our choice of
momenta,\ we expect the $\cR^{\RI}_{ij}(m_q,p,g_0)$ to be functions of $p^2$
to good approximation. Therefore, the leading terms in the OPEs of these
ratios at finite $m_q$, including discretization error terms, are of the form
(see also \cite{Garron:2003cb}):
\be
\cR^{\RI}_{ij}(m_q,p,g_0)=\cdots+\frac{A_{ij}^{(2)}(m_q,g_0,p^2)}{p^2}+
U^{\RI}_{ik}(p^2)\,
Z^{\RGI}_{kj}(g_0)
+B_{ij}^{(2)}(g_0,p^2)\, (ap)^2+\cdots\ ,
\labell{eq:rriope}
\ee
where the ellipses stand for higher-order terms in the OPE and
sub-leading discretization errors.  Here $U^{\RI}_{ik}(p^2)$ describes
the running of the renormalization constants in the RI/MOM scheme,
$Z_{ij}^{\RI}(p^2)=U^{\RI}_{ik}(p^2)Z^{\RGI}_{kj}$, where the
superscript ``RGI'' (renormalization group invariant) indicates that the
corresponding renormalization constant is the constant which
translates lattice results into scale- and scheme-independent
quantities.~\footnote{To specify our normalization conventions for RGI
quantities, we give as an example the factor $[U^{\RI}]_{11}$ at one
loop. It is $(2\beta_0\alpha_s/4\pi)^{4/2\beta_0}$ where, of course,
$\beta_0=11-2N_f/3$ is the coefficient of the one loop beta
functions.} The $p^2$ dependence in $A_{ij}^{(2)}$ and $B_{ij}^{(2)}$
is specified to accommodate the expected logarithmic corrections to
the power behavior in $p^2$. Note that $B_{ij}^{(2)}$ is independent
of quark mass, because this dependence only appears at higher orders
in the OPE.
  
To determine whether power corrections and discretization errors are relevant
in extracting the renormalization constants from the
$\cR^{\RI}_{ij}(m_q,p,g_0)$, we consider the ``RGI'' ratios:
\be
\cR^{\RGI}_{ij}(m_q,p,g_0)=[U^{\RI}]^{-1}_{ik}(p^2)\cR^{\RI}_{kj}(m_q,p,g_0)
\ee
obtained from the $\cR^{\RI}_{ij}(m_q,p,g_0)$ by dividing out the
running.  In the absence of discretization errors and sub-leading
perturbative and OPE terms, these ``RGI'' ratios should be constant
functions of $p^2$. In \fig{fig:Z11RGI600vsp2}, we plot
$\cR^{\RGI}_{ij}(m_q,p,g_0)$ for $i=j=1$ as functions of $p^2$ for our
smallest and largest quarks masses, $am_q=0.03$ and $0.10$ at
$\beta=6.0$. We do the same in \fig{fig:Z23RGI600vsp2} for
$\cR^{\RGI}_{ij}(m_q,p,g_0)$ with $i,\,j\in {2,3}$, and in
\fig{fig:Z45RGI600vsp2} for the mixing of operators $O_4$ and
$O_5$. Here the running is implemented at two loops
\cite{Ciuchini:1995cd,Buras:2000if}, with $N_f=0$ and
$\Lambda_\mathrm{QCD}=0.238(19)\,\gev$ from \cite{Capitani:1998mq}.  A
number of observations are in order:
\begin{itemize}
\item to a very good approximation, $\cR^{\RGI}_{ij}(m_q,p,g_0)$ are functions
  of $p^2$;
\item the $\cR^{\RGI}_{ij}(m_q,p,g_0)$ are not generically constant
functions of $p^2$ at values of $p^2$ where one might consider
extracting the renormalization constants; when they are, this behavior
seems to be the result of a turnover instead of genuine
$p^2$-independence;
\item the $\cR^{\RGI}_{ij}(m_q,p,g_0)$ do in some cases exhibit a significant
  mass dependence, but that dependence tends to vanish when $p^2$ increases as
  predicted by the OPE;
\item discretization errors are clearly visible but remain moderate and there
  is no evidence for contributions of order higher than $(ap)^2$.
\end{itemize}

\begin{figure}[t]
\begin{center}
\psfig{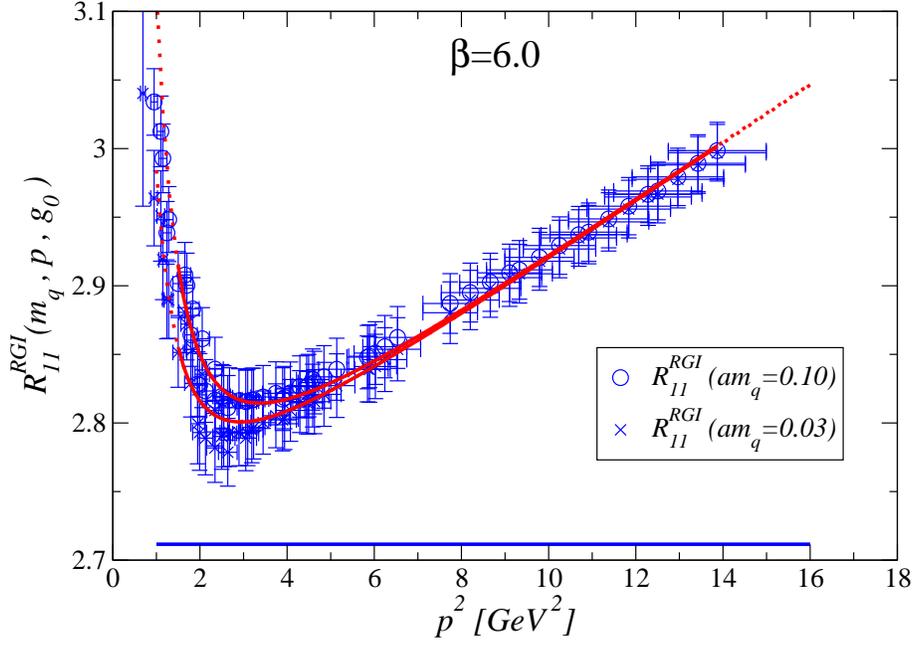}
\caption{\labell{fig:Z11RGI600vsp2} Plot of the renormalization ratio
  $\cR^{\RGI}_{11}(m_q,p,g_0)$ for the standard model operator $O_1$ as a
  function of four-momentum squared, $p^2$, for our most massive ($am_q=0.10$,
  circle) and our lightest ($am_q=0.03$, cross) quarks.  The solid curve for
  each data set illustrates the combined fit to the parameterization of
  \eq{eq:rriope}, with the terms indicated in \tab{tab:ritermslr}, and in
  range fitted, as described in the text. The dotted curve is an extension of
  this fit beyond the fit range. The solid horizontal line is the resulting
  value of $Z_{11}^\RGI(g_0)$.}
\end{center}
\end{figure}

\begin{figure}[t]
\begin{center}
\psfig{file=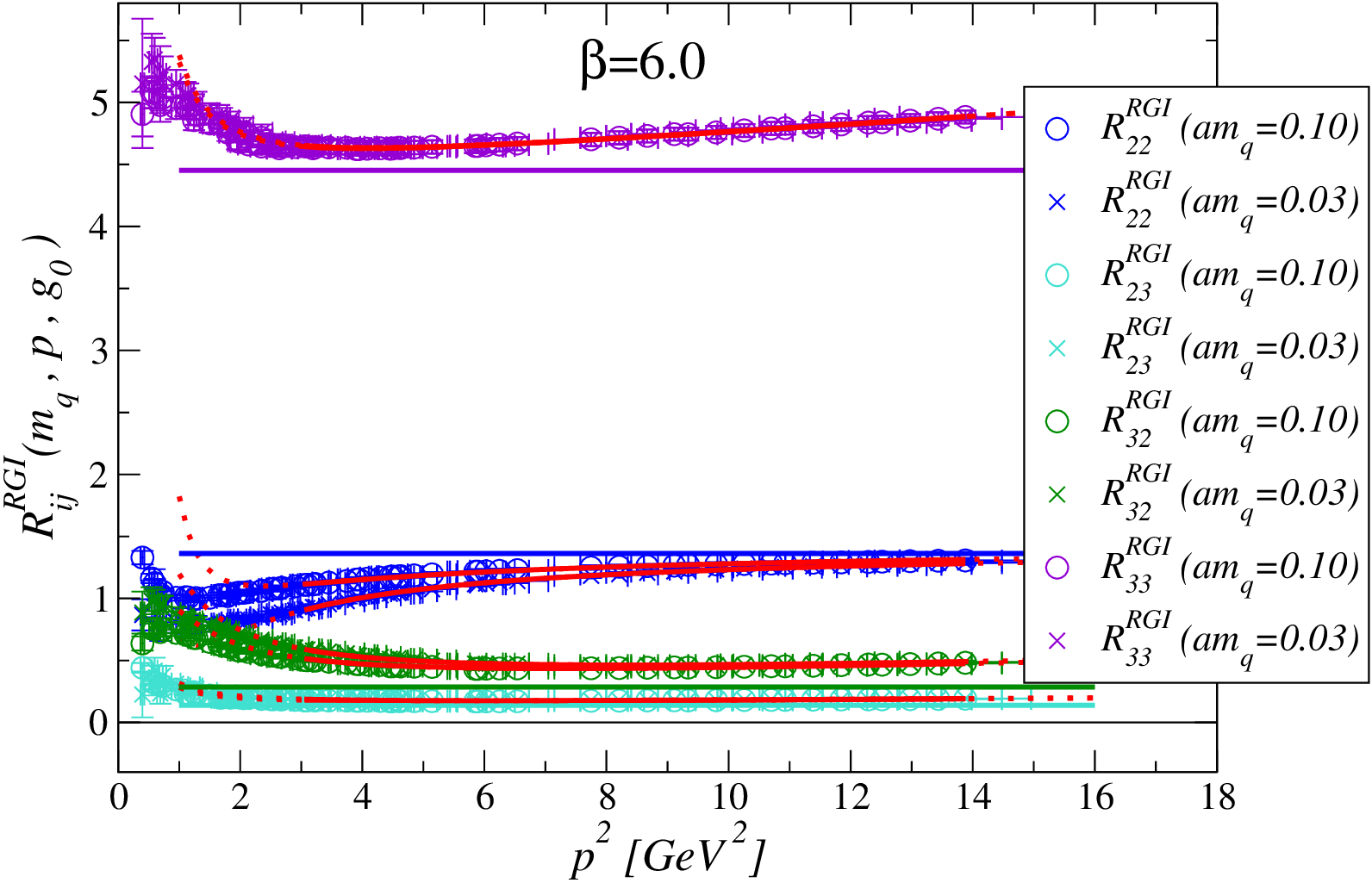,width=12cm}
\caption{\labell{fig:Z23RGI600vsp2} Same as \fig{fig:Z11RGI600vsp2}, but 
  for the operator pair $\{O_2,\,O_3\}$. Here all eight ratios are fit
  simultaneously.}
\end{center}
\end{figure}

\begin{figure}[t]
\begin{center}
\psfig{file=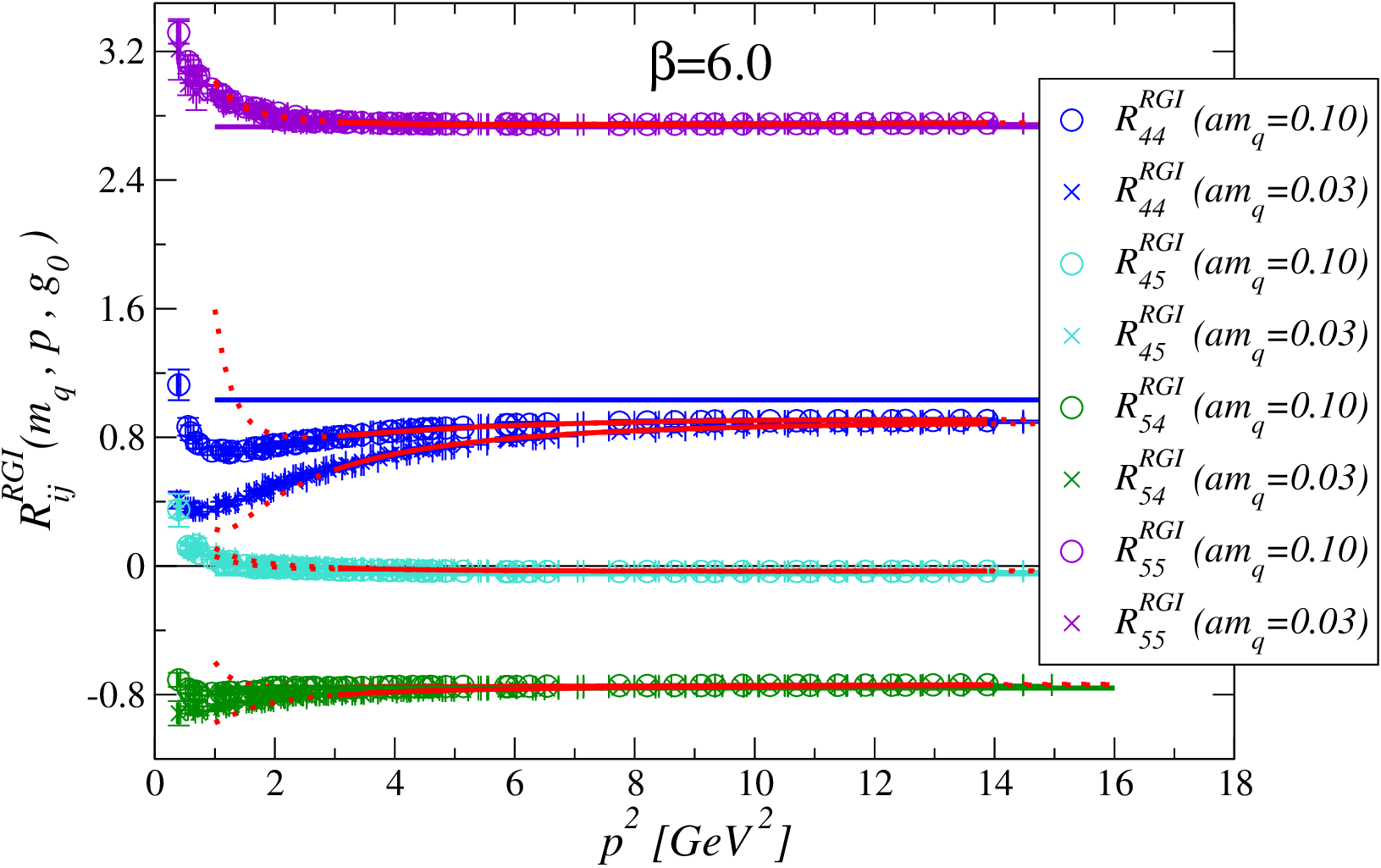,width=12cm}
\caption{\labell{fig:Z45RGI600vsp2} Same as \fig{fig:Z11RGI600vsp2}, but 
  for the operator pair $\{O_4,\,O_5\}$. Here all eight ratios are fit
  simultaneously.}
\end{center}
\end{figure}

Our conclusion is that power corrections and discretization errors may give
significant contributions and they should be accounted for in extracting the
renormalization constants, as already briefly discussed in
\cite{Garron:2003cb}. This is even more true at $\beta=5.85$ where
discretization errors enter at lower physical momenta.  So,
instead of performing, in some cases delicate, chiral extrapolations
of $\cR^{\RI}_{ij}(m_q,p,g_0)$ to $m_q=0$~\footnote{We have performed
the calculation at five quark masses which are not that close to the
chiral limit, and the full mass dependence of these functions,
especially in the quenched approximation, is not known.} and taking
$Z_{ij}^{\RI}(g_0,p^2)=
\cR^{\RI}_{ij}(m_q,p,g_0)$ for $p^2\simeq 4\,\gev^2$ as is usually
done \cite{Martinelli:1995ty,Donini:1999sf}, we
fit the $\cR^{\RI}_{ij}(m_q,p,g_0)$ at finite $m_q$ to OPE forms such
as the one given in \eq{eq:rriope}. This procedure not only eliminates
the mass effects through their $p^2$ dependence, but also other
power-suppressed contributions as well as discretization errors. In
performing these fits, we neglect possible logarithmic corrections in
$p^2$ in the coefficients $A_{ij}^{(2,4)}$ and $B_{ij}^{(2)}$. They
are at least subdominant to the leading corrections and
sub-subdominant compared to the perturbative behavior that we want to
isolate.~\footnote{$1/p^4$ corrections are also formally
sub-subdominant. However our results often clearly indicate the
presence of such terms.} Moreover, they are not known analytically and
they would be difficult to identify in the data. We also neglect
discretization errors in $(am)^2$ since they should be significantly
smaller than either the $(ap)^2$ discretization errors or the $1/p^2$
corrections in the range of $p^2$ considered.

We obtain our central values for the renormalization constants by fitting to
all points $p^2$ such that $3\,\gev^2\le p^2\le 14\,\gev^2$ at
$\beta=6.0$ and $3\,\gev^2\le p^2\le 8.5\,\gev^2$ at
$\beta=5.85$. The reason for
this rather large range of $p^2$ is that we want to make sure that all OPE and
discretization terms, which make a significant contribution to
$\cR^{\RI}_{ij}(m_q,p,g_0)$ in regions where we can hope to extract reliable
values of the renormalization constants, are accurately measured with the
data. Moreover, down to $3\,\gev^2$, the two-loop running remains under
control. For $\cR^{\RI}_{11}(m_q,p,g_0)$, we consider even lower momenta, down
to $p^2=1.5\,\gev^2$. The reason for this is the presence of a large power
correction whose coefficient can only be determined at these low values of
$p^2$. And because the running of $O_1$ is mild, the two-loop evolution of
this operator remains sensible even down to these low values of the momentum.
The terms kept in the fits for the various renormalization constants are
summarized in \tab{tab:ritermslr}.

\begin{table}
\begin{center}
\caption{\labell{tab:ritermslr} Terms retained 
  in the fits of the renormalization ratios
  $\cR^{\RI}_{ij}(m_q,p,g_0)$ to the OPE form of \eq{eq:rriope}, in the
  range $3\,\gev^2\le p^2\le 14\,\gev^2$ at $\beta=6.0$ and
  $3\,\gev^2\le p^2\le 8.5\,\gev^2$ at $\beta=5.85$. The
  discretization term in $(ap)^2$ is treated as mass independent, as
  discussed in the text. At $\beta=6.0$,
  $(am_q^{min},am_q^{max})=(0.030,0.100)$ and
  $(am_q^{min},am_q^{max})=(0.040,0.132)$ at $\beta=5.85$. $\cR_S^\RI$ is
  the ratio from which the renormalization constant of the scalar
  density is obtained. It is fit to an OPE form starting from
  $2\,\gev$.}
\begin{tabular}{cccccc}
\hline
\hline
& \multicolumn{2}{c}{$am_q^{min}$} & \multicolumn{2}{c}{$am_q^{max}$} & \\
\hline 
& $1/p^4$ & $1/p^2$
& $1/p^4$ & $1/p^2$ & $(ap)^2$\\
\hline
$\cR^\RI_{11}$ & $\times$ & & $\times$ & & $\times$ \\
$\cR^\RI_{22}$ & $\times$ & $\times$ & $\times$ & $\times$ & $\times$ \\
$\cR^\RI_{23}$ & & $\times$ & & $\times$ & $\times$ \\
$\cR^\RI_{32}$ & & $\times$ & & $\times$ & $\times$ \\
$\cR^\RI_{33}$ & $\times$ & & $\times$ & & $\times$ \\
$\cR^\RI_{44}$ & $\times$ & $\times$ & $\times$ & $\times$ & $\times$ \\
$\cR^\RI_{45}$ & & $\times$ & & $\times$ & $\times$ \\
$\cR^\RI_{54}$ & $\times$ & $\times$ & $\times$ & $\times$ & $\times$ \\
$\cR^\RI_{55}$ & $\times$ & & $\times$ & & $\times$ \\
$\cR^\RI_S$ & $\times$ & $\times$ & $\times$ & $\times$ & $\times$ \\
\hline
\hline
\end{tabular}
\end{center}
\end{table}

We initially individually fit $\cR^{\RI}_{ij}(m_q,p,g_0)$ obtained for
our lightest or next to lightest and heaviest quark masses
($(am_q^{min},am_q^{max})=(0.030,0.100)$ at $\beta=6.0$ and
$(0.040,0.132)$ at $\beta=5.85$). We find that, despite a factor of
over 3 between these two masses, the resulting $Z^{\RGI}_{ij}(g_0)$,
as well as the discretization error coefficients $B_{ij}(g_0)$, are
independent of quark mass within statistical errors, giving us
confidence in the consistency of our OPE description. Therefore, for
our final results we perform a combined fit to the
$\cR^{\RI}_{ij}(m_q,p,g_0)$ for the two masses, constraining
$Z^{\RGI}_{ij}(g_0)$ and $B_{ij}(g_0)$ to be mass-independent, but
allowing, of course, $A_{ij}^{(2)}$ to be different for the two
masses. The resulting fits at $\beta=6.0$ are shown in
\figs{fig:Z11RGI600vsp2}{fig:Z45RGI600vsp2}, together with the
horizontal line describing the behavior of the perturbative term
alone, as reconstructed from the fit. The distance between this line
and the data points quantifies the impact of OPE and discretization
corrections to the leading perturbative term. Similar plots are
obtained at $\beta=5.85$, though the stability of the fits is
reduced by the presence of discretization errors at even lower values
of $p^2$. The resulting renormalization constants,
$Z^\RI_{ij}(2\,\gev)$, in the RI/MOM scheme at $2\,\gev$ are
summarized in \tab{tab:ZOijRI600lr} for $\beta=6.0$ and
\tab{tab:ZOijRI585lr} for $\beta=5.85$. Since the $\msbar$-NDR scheme
is also often used, for completeness we give the renormalization
constants in that scheme in the \app{sec:appndrrenormcsts} in
\tabs{tab:ZOijNDR600lr}{tab:ZOijNDR585lr}.

To renormalize the $B$-parameters, we further need to renormalize the
matrix elements' vacuum saturation values. For $O_{2,\cdots,5}$, these
are given by matrix elements of the pseudoscalar density, so that the
renormalization constant needed is $Z_P$. This constant is actually
difficult to obtain directly in the RI/MOM scheme of
\cite{Martinelli:1995ty}, because the relevant vertex receives
chirally enhanced power corrections
\cite{Martinelli:1995ty}. Fortunately, the chiral symmetry of
Neuberger fermions guarantees that $Z_P=Z_S$, where $Z_S$ is the
renormalization constant for the scalar density, which is free of
Goldstone pole contaminations. Thus, we consider a ratio
$\cR_S^\RI(m_q,p,g_0)$, which is analogous to the ratios
$\cR_{ij}^\RI(m_q,p,g_0)$ defined in \eq{eq:rirats}, with the
operators $O_i$ and $O_j$ replaced by the scalar density and the
squares replaced by first powers in the numerator. As above, we study
the momentum dependence of this ratio and fit it to an OPE expression
of the form given in \eq{eq:rriope}, in the interval $2\,\gev^2\le
p^2\le 14\,\gev^2$ at $\beta=6.0$ and $2\,\gev^2\le p^2\le
8.5\,\gev^2$ at $\beta=5.85$. We find that a good description of the
data is obtained by retaining the discretization term proportional to
$(ap)^2$ together with $1/p^2$ and $1/p^4$ OPE terms (see
\tab{tab:ritermslr}). The data and the fit at $\beta=6.0$ are shown in
\fig{fig:ZSRGI600vsp2}, together with the horizontal line describing
the behavior of the perturbative term alone, as reconstructed from the
fit. One difference here is that the running of the operator is
implemented at four loops
\cite{Chetyrkin:1999pq,vanRitbergen:1997va,Vermaseren:1997fq}, with
$N_f=0$ and the same value of $\Lambda_\mathrm{QCD}$ as above. The
result in the RI/MOM scheme at $2\,\gev$ is given in \tab{tab:zsri}.
The value at the same scale in the $\msbar$ scheme is given in
\app{sec:appndrrenormcsts} in \tab{tab:zsmsbar}.  These constants are
also required for determining the renormalized quark masses of
\tabs{tab:msplusmhatRI600and585}{tab:msplusmhatMSbar600and585}, from
the corresponding bare quark masses.

\begin{figure}[t]
\begin{center}
\psfig{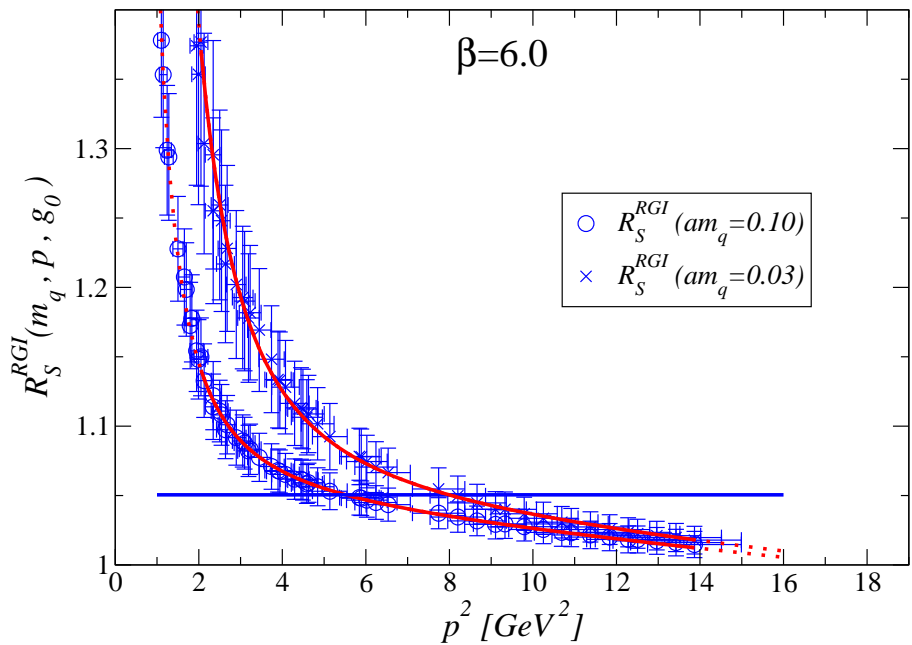}
\caption{\labell{fig:ZSRGI600vsp2} Same as \fig{fig:Z11RGI600vsp2}, but 
  for the scalar density. Here, running is implemented at four loops, as
  described in the text.}
\end{center}
\end{figure}

Though it is straightforward to obtain the renormalization constants
for the $B$-parameters from those for the operators in
\tabs{tab:ZOijRI600lr}{tab:ZOijRI585lr} and for $Z_P$ in \tab{tab:zsri}, we
nevertheless give our results for these constants in the RI/MOM scheme
at $\beta=6.0$ in
\tab{tab:ZBijRI600lr} and at $\beta=5.85$ in
\tab{tab:ZBijRI585lr}, because they may be of use
and because we can account for the correlations in the ratios of
$Z_{ij}$ to $Z_P^2$.  Results in the $\msbar$-NDR scheme are
summarized in \tabs{tab:ZBijNDR600lr}{tab:ZBijNDR585lr} in
\app{sec:appndrrenormcsts}.

\medskip

To estimate the systematic errors associated with our procedure for
determining renormalization constants, we consider fits to a higher
momentum range, where power corrections in $1/p^2$ will play a less
important r\^ole.  We choose the range $7.5\,\gev^2\le p^2\le
14\,\gev^2$ at $\beta=6.0$ and $5\,\gev^2\le p^2\le 8.5\,\gev^2$ at
$\beta=5.85$, where momenta belong to the high momentum cone described
below \eq{eq:rirats}. We also eliminate the power corrections in
$1/p^2$ to which our results are not sensitive in this higher range of
momenta. The list of terms kept is summarized in \tab{tab:ritermssr}
and the resulting renormalization constants are given in
\tabs{tab:ZOijRI600sr}{tab:ZOijRI585sr} in the RI/MOM scheme. The
resulting values for $Z_S$ are given in \tab{tab:zsri} and the results
for the renormalization constants of the $B$-parameters are summarized
in \tabs{tab:ZBijRI600sr}{tab:ZBijRI585sr}.

\begin{table}
\begin{center}
\caption{\labell{tab:ritermssr} Same as \tab{tab:ritermslr},
but for the restricted range $7.5\,\gev^2\le p^2\le
  14\,\gev^2$ at $\beta=6.0$ and $5\,\gev^2\le p^2\le 8.5\,\gev^2$ at
  $\beta=5.85$.}
\begin{tabular}{ccccccc}
\hline
\hline
& \multicolumn{2}{c}{$am_q^{min}$} & \multicolumn{2}{c}{$am_q^{max}$} \\
\hline 
& $1/p^4$ & $1/p^2$
& $1/p^4$ & $1/p^2$ & $(ap)^2$\\
\hline
$\cR^\RI_{11}$ & & & & & $\times$ \\
$\cR^\RI_{22}$ & & $\times$ & & $\times$ & $\times$ \\
$\cR^\RI_{23}$ & & $\times$ & & $\times$ & $\times$ \\
$\cR^\RI_{32}$ & & $\times$ & & $\times$ & $\times$ \\
$\cR^\RI_{33}$ & & & & & $\times$ \\
$\cR^\RI_{44}$ & & $\times$ & & $\times$ & $\times$ \\
$\cR^\RI_{45}$ & & $\times$ & & $\times$ & $\times$ \\
$\cR^\RI_{54}$ & & $\times$ & & $\times$ & $\times$ \\
$\cR^\RI_{55}$ & & & & & $\times$ \\
$\cR^\RI_S$ & & $\times$ & $\times$ & & $\times$ \\
\hline
\hline
\end{tabular}
\end{center}
\end{table}

As the renormalization constants obtained from the two fitting ranges
indicate, our description of the RI/MOM Green functions is very stable
at $\beta=6.0$, giving us confidence that we have isolated correctly
the various OPE and discretization terms which contribute to these
functions. However it is clear that for a high precision calculation
of matrix elements, a non-perturbative renormalization procedure in
which corrections in powers of $1/p^2$ are absent and in which
perturbation theory is only used at much higher scales would be highly
desirable, especially on coarser lattices. Proposals in that direction
have been made very recently in
\cite{Luscher:2006df,Taniguchi:2006qw}.

\section{Mass dependence and physical matrix elements}

\labell{sec:massdep}

Having determined the renormalization constants, we can now
renormalize the bare matrix elements and $B$-parameters for our
various meson masses.  Renormalizing these quantities before studying
their behavior in mass is useful because it allows for comparisons of
this behavior with results obtained in other simulations, performed
with either different actions or different parameters. In our case, we
undertake such a comparison between the results obtained on our finer
and coarser lattices, at $\beta=6.0$ and $\beta=5.85$,
respectively. Moreover, this renormalization is also important if one
is to perform a combined chiral and continuum extrapolation using
finite lattice spacing chiral perturbation theory.

In \tabs{tab:birivsam600}{tab:birivsam585} we present the results
obtained at $\beta=6.0$ and $\beta=5.85$, respectively, for the
$B$-parameters in the RI/MOM scheme at $2\,\gev$, as functions of bare
quark mass. The corresponding results for the renormalized BSM ratios,
$R_i^\BSM$, are given in \tab{tab:ririvsam600} for $\beta=6.0$ and
\tab{tab:ririvsam585} for $\beta=5.85$. In \tab{tab:girivsam600} for
$\beta=6.0$ and \tab{tab:girivsam585} for $\beta=5.85$, we summarize
our results for the couplings $G_i$ of \eq{eq:gidef}, again as a
function of light-quark mass.  Finally,
\tabs{tab:dirivsam600}{tab:dirivsam585} display the results that we
obtain for $D_{7,8}^{3/2}$ in the RI/MOM scheme at $\beta=6.0$ and
$\beta=5.85$, respectively. For completeness, we present all of these
results renormalized in the $\msbar$-NDR scheme at $2\,\gev$ in
\app{sec:appmassdepndr}, in \tabs{tab:bindrvsam600}{tab:dindrvsam585}.

In order to finalize our calculation of the $B$-parameters, BSM ratios
$R_i^\mathrm{BSM}$ and couplings $G_i$ relevant for $K^0$-$\bar K^0$
mixing beyond the standard model, we have to interpolate our
renormalized results to the kaon mass.  To perform the interpolation
we fit our data, with the exclusion of the single point for
$am_{q}=0.030$ at $\beta=5.85$ which we leave out for the reasons
explained in \protect\sec{sec:2ptfns}, to simple polynomial forms in
$M^2/(4\pi F)^2$. As we did for setting the scale in \sec{sec:2ptfns},
we fix the kaon point to be $M^2/(4\pi F)^2=M_K^2/(4\pi F_K)^2$.  More
complicated chiral perturbation theory fits could be tried, but their
applicability in our mass range is questionable and for the simple
interpolations that we have to perform, any reasonable description of
the data is sufficient.

We begin with the interpolation of the $B$-parameters. We find that
$B_1$, and $B_5$ are consistent with linear behavior in $M^2/(4\pi
F)^2$ while $B_2$, $B_3$ and $B_4$ are well fit by parabolas in our
range of masses. We show these fits for illustrative purposes in
\fig{fig:biriinter600and585}, but do not reproduce the fit parameters
here since they are not of physical interest. Our results for the
$B$-parameters at the kaon mass in the RI/MOM scheme at $2\,\gev$ are
summarized in \tab{tab:riresatmk600and585} at $\beta=6.0$ and 5.85.

\begin{figure}[t]
\begin{center}
\psfig{file=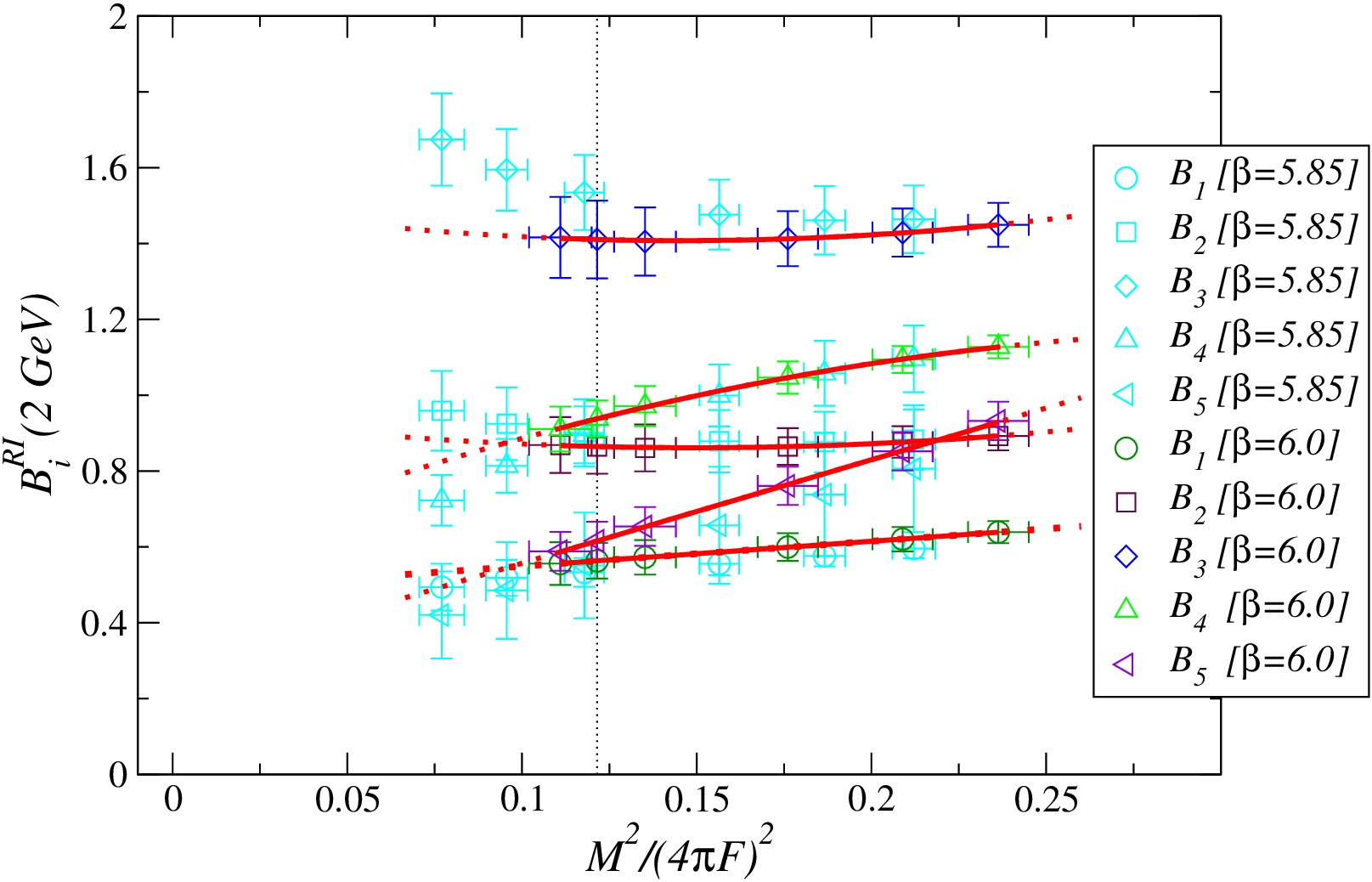,width=12cm}
\caption{\labell{fig:biriinter600and585} Mass-dependence, in terms of
  the variable $M^2/(4\pi F)^2$, of the $B$-parameters $B_i$,
  $i=1,\cdots,5$, in the RI/MOM scheme at $2\,\gev$. The solid curves
  are the results of the fits described in the text, and are plotted
  in the fit region. The fits are used to interpolate the results to
  the kaon point $M^2/(4\pi F)^2=M_K^2/(4\pi F_K)^2$, shown as a
  vertical dotted line.  The dashed curves are an extension of the fit
  curves outside the fit range.}
\end{center}
\end{figure}

We now turn to the BSM ratios $R^\mathrm{BSM}_i$. Here we find that
$R^\mathrm{BSM}_4$ and $R^\mathrm{BSM}_5$ are well fit by a straight
line while $R^\mathrm{BSM}_2$ and $R^\mathrm{BSM}_3$ require quadratic
terms in $M^2/(4\pi F)^2$. The results of the interpolations to the
kaon point are reproduced in \tab{tab:riresatmk600and585}
and are illustrated in \fig{fig:riinter600and585}.

\begin{figure}[t]
\begin{center}
\psfig{file=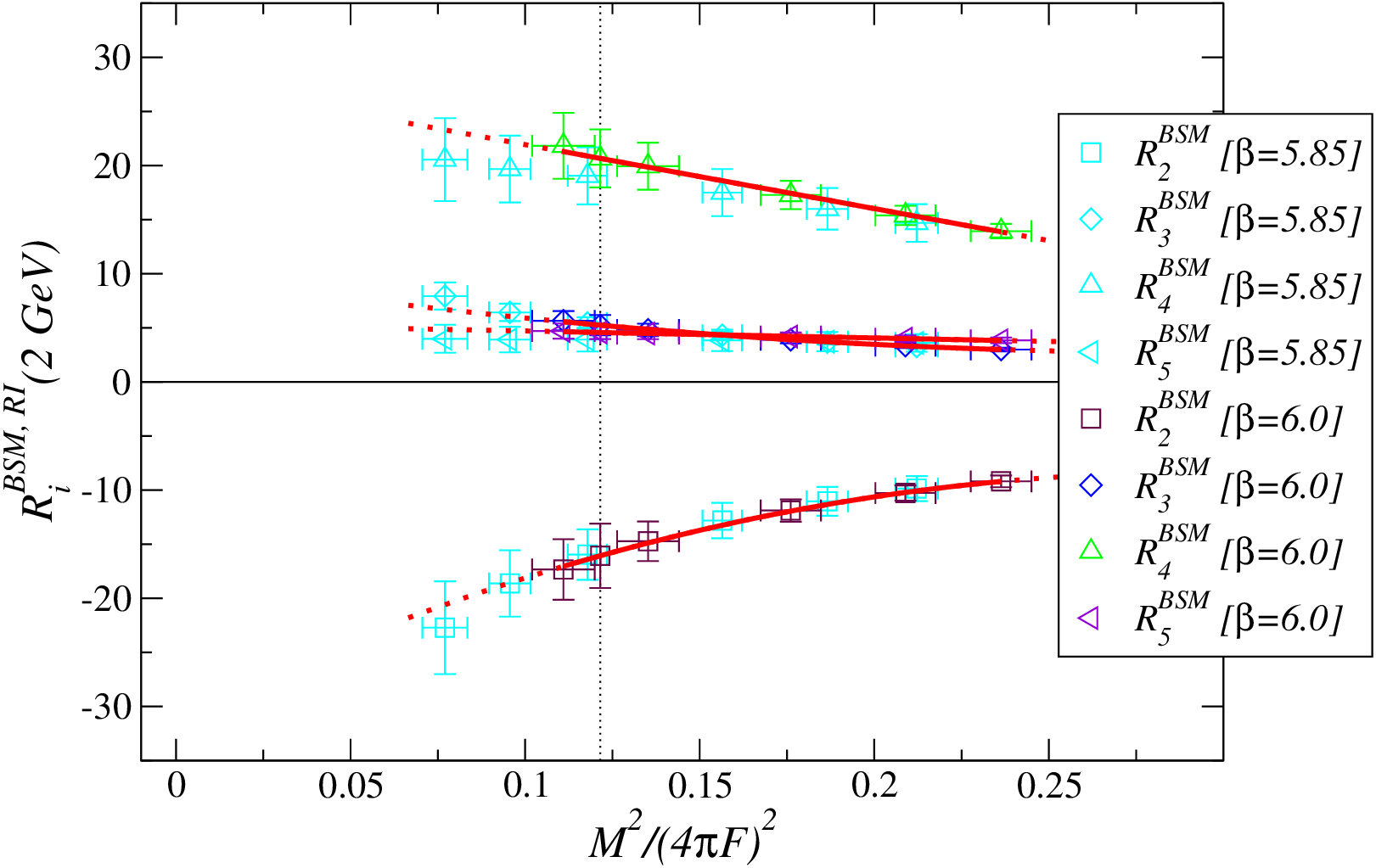,width=12cm}
\caption{\labell{fig:riinter600and585} Same as
  \protect\fig{fig:biriinter600and585}, but for the BSM ratios
  $R_i^\BSM$, $i=2,\cdots,5$.}
\end{center}
\end{figure}

We further interpolate the couplings $G_i$. As \fig{fig:giinter600}
confirms, $G_5$ is well fit by a constant, $G_4$ by a straight line,
while $G_1$, $G_2$ and $G_3$ require quadratic terms in $M^2/(4\pi
F)^2$. Again, we do not reproduce the fit parameters here. We only
note that in the chiral limit $G_1$ is consistent with zero, which is
the expected behavior. The results for the couplings $G_i$ at the kaon
mass in the RI/MOM scheme at $2\,\gev$ are given in
\tab{tab:riresatmk600and585}.

\begin{figure}[t]
\begin{center}
\psfig{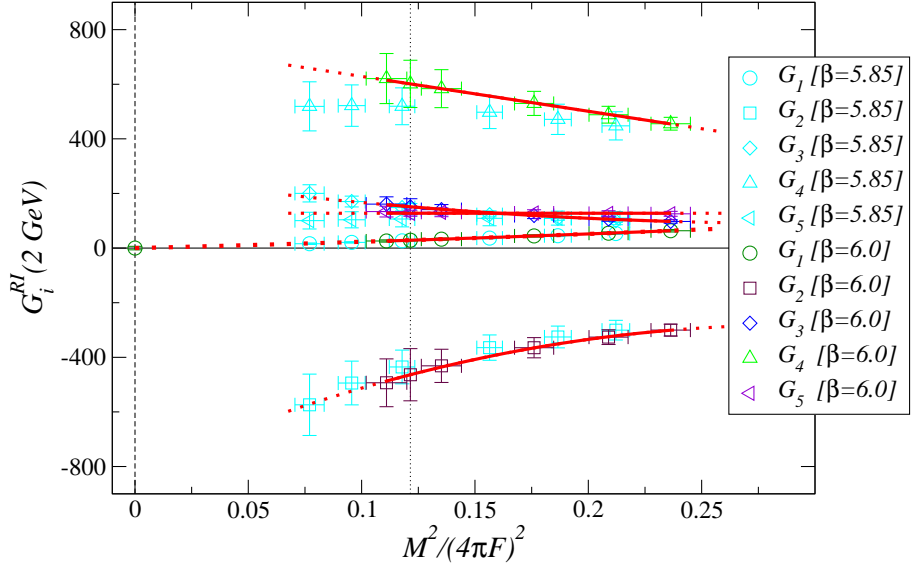}
\caption{\labell{fig:giinter600} Same as
  \protect\fig{fig:biriinter600and585}, but for the ``couplings''
  $G_i$, $i=1,\cdots,5$. Also shown, for indicative purposes, is an
  extrapolation of $G_1$ to the chiral limit, which is denoted by a
  dashed vertical line at $M^2/(4\pi F)^2=0$.}
\end{center}
\end{figure}

\begin{table}[t]
\begin{center}
\caption{\labell{tab:riresatmk600and585} Results at the kaon mass in
the RI/MOM scheme at $2\,\gev$ for $\beta=6.0$ and $\beta=5.85$. The
first error is statistical and the second originates from the
systematic uncertainty in the renormalization constants.}
\begin{tabular}{ccccccc}
\hline
\hline
& \multicolumn{3}{c}{$\beta=6.0$} & \multicolumn{3}{c}{$\beta=5.85$}\\
\hline
$i$ & $B_i$ & $G_i$ & $R_i$ & $B_i$ & $G_i$ & $R_i$  \\
\hline
1 & $0.563(47)(7)$ & $28.8(25)(3)$ & $1$ & $0.534(36)(4)$ & $27.3(18)(0)$ & $1$ \\ 
2 & $0.865(72)(9)$ & $-464.(96)(11)$ & $-16.1(30)(6)$ & $0.898(89)(44)$ & $-430.(69)(40)$ & $-15.8(25)(14)$ \\ 
3 & $1.41(10)(6)$ & $151.(30)(4)$ & $5.24(93)(9)$ & $1.53(10)(40)$ & $146.(17)(13)$ & $5.37(64)(51)$ \\ 
4 & $0.938(48)(13)$ & $601.(87)(0)$ & $20.7(27)(2)$ & $0.904(73)(106)$ & $514.(69)(24)$ & $18.8(27)(7)$ \\ 
5 & $0.616(51)(14)$ & $128.(11)(1)$ & $4.57(60)(1)$ & $0.56(14)(1)$ & $108.(27)(17)$ & $3.9(11)(6)$ \\ 
\hline
\hline
\end{tabular}
\end{center}
\end{table}

From these results for $B_i$, $G_i$ and $R^\BSM_i$ at the kaon mass in
the RI/MOM scheme, it is straightforward to obtain the corresponding
results in the $\msbar$-NDR scheme at $2\,\gev$. This is done in
\app{sec:appndrres} and the results are presented there in
\tab{tab:ndrresatmk600and585}.

\medskip

Finally, we turn to the most delicate part of this discussion: the chiral
extrapolation of the electroweak penguin matrix elements. As stated in the
Introduction, the quantities $D_{7,8}^{3/2}$ of \eq{eq:d78def} have the same
chiral logarithm at NLO in the quenched and $N_f=3$ theories, while the
corresponding matrix elements or parameters $G_{5,4}$ do not. Indeed, at NLO
in the $N_f=3$ theory \cite{Gasser:1984gg},
\be
F=F_\chi\l(1-\frac32\frac{M^2}{(4\pi F)^2}\ln 
\frac{M^2}{(4\pi \rho_F F)^2}+O(p^4)\r)
\ee
and \cite{Becirevic:2004qd}
\be
\la \pi^+|Q_{7,8}^{3/2}|K^+\ra=\mathcal{M}^{3/2,\chi}_{7,8}
\l(1-4\frac{M^2}{(4\pi F)^2}\ln 
\frac{M^2}{(4\pi\rho_{7,8}^{3/2,\mathrm{M}}F)^2}+O(p^4)\r)
\ .\ee
In the quenched theory, we have at NLO \cite{Bernard:1992mk},
\be
F=F_\chi\l(1+\frac{M^2}{(4\pi F)^2}\ln\frac{1}{\rho_F^2}+O(p^4)\r)
\ee
and \cite{Villadoro:privcomm04}
\be
\la \pi^+|Q_{7,8}^{3/2}|K^+\ra=\mathcal{M}^{3/2,\chi}_{7,8}
\l(1-\frac{M^2}{(4\pi F)^2}\ln 
\frac{M^2}{(4\pi\rho_{7,8}^{3/2,\mathrm{M}}F)^2}+O(p^4)\r)
\ ,\ee
where it is understood that the low energy constants $\rho_F$, $\ldots$ may
differ in the quenched and $N_f=3$ theories. Thus, for $D_{7,8}^{3/2}$, we
find in both cases, the following chiral behavior:
\be
D_{7,8}^{3/2}=D^{3/2,\chi}_{7,8}
\l(1-\frac{M^2}{(4\pi F)^2}\ln 
\frac{M^2}{(4\pi\rho_{7,8}^{3/2,D} F)^2}+O(p^4)\r)
\labell{eq:d78chilog}
\ ,\ee
where again, the low energy constants are distinct in the two
theories.  However, the fact that the functional behavior is the same
and the fact that the coefficient of the logarithm is small suggests
that $D_{7,8}^{3/2,\chi}$ may be comparable in the two theories,
especially since our lattice has been tuned to reproduce results at
the kaon mass and the extrapolation should be mild. This mildness of
the mass behavior is actually seen in our results for $D_{7,8}^{3/2}$
renormalized in the RI/MOM scheme at $2\,\gev$, which are plotted as
functions of $M^2/(4\pi F)^2$ in \fig{fig:d78inter600}.

\begin{figure}[t]
\begin{center}
\psfig{file=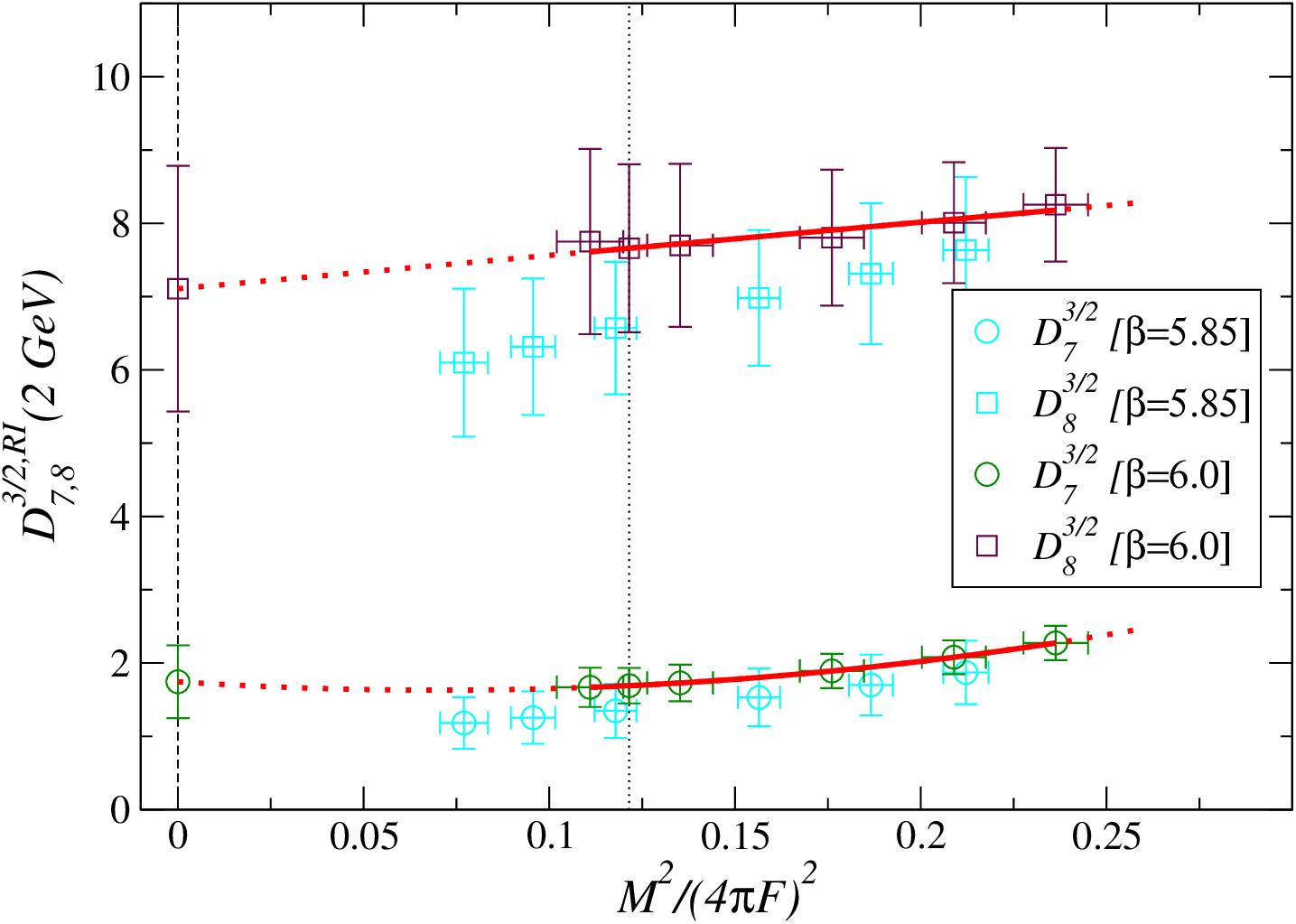,width=12cm}
\caption{\labell{fig:d78inter600} Mass-dependence, in terms of the variable 
  $M^2/(4\pi F)^2$, of the ratios $D_{7,8}^{3/2}$ in the RI/MOM scheme at
  $2\,\gev$. The solid curves are the results of the fits described in the
  text, and are plotted in the fit region. The fits are used to interpolate
  the results to the kaon point $M^2/(4\pi F)^2=M_K^2/(4\pi F_K)^2$, shown as
  a vertical dotted line. The dashed curves are an extension of the fit curves
  outside the fit range. Also shown, for indicative purposes, are
  extrapolations of $D_{7,8}^{3/2}$ to the chiral limit, which is denoted by a
  dashed vertical line at $M^2/(4\pi F)^2=0$.}
\end{center}
\end{figure}

Instead of attempting a hazardous chiral extrapolation of
$D_{7,8}^{3/2}$, we choose to interpolate these quantities to the kaon
point. Because $F_K$ is used to fix the lattice cutoff, we take
$D_{7,8}^{3/2}(\mu,M_K^2)\equiv F_K^2\times G_{5,4}(\mu,M_K^2)$, with
$G_{5,4}(\mu,M_K^2)$ obtained above in the RI/MOM scheme at
$2\,\gev$. We then use the fact that any reasonable chiral
extrapolation (linear, quadratic, log form of \eq{eq:d78chilog}, log
form plus a quadratic term in $M^2/(4\pi F)^2$, $\ldots$) gives values
which are compatible with the results at the kaon point. This is shown
in \fig{fig:d78inter600} where $D_8^{3/2}$ is extrapolated linearly
and $D_8^{3/2}$ quadratically for illustration. Thus, the systematic
uncertainty due to the chiral extrapolation should be smaller than our
statistical errors and we have approximately $D_{7,8}^{3/2,\chi}\simeq
D_{7,8}^{3/2}(M_K^2)$.  Now, for determining from $D_{7,8}^{3/2,\chi}$
the chiral limit value of the $K\to\pi\pi$ matrix element required to
calculate $\epsilon'$, we can use \eq{eq:kpipiproptokpi}, taking $\la
(\pi\pi)_{I=2}|Q_{7,8}|K^0\ra_\chi\propto F_\chi D_{7,8}^{3/2,\chi}$,
with $F_\chi=87(5)\,\mev$ \cite{Gasser:1984gg}.  Alternatively, since
we are working to leading chiral order, we can choose to remain at the
kaon point and consider $\la (\pi\pi)_{I=2}|Q_{7,8}|K^0\ra_\chi\propto
F_K D_{7,8}^{3/2}(M_K^2)$. For our final value of the matrix elements,
we take an average of the two central values and use their spread as
an estimate of the chiral extrapolation error.  This leads to a $\pm
15\%$ systematic error on $\la
(\pi\pi)_{I=2}|Q_{7,8}|K^0\ra_\chi$. \tab{tab:q7q8RI600and585} summarizes
all of these results in the RI/MOM scheme at $2\,\gev$ for
$\beta=6.0$ and 5.85. The results in the $\msbar$-NDR scheme are
reproduced in \tab{tab:q7q8NDR600and585} of \app{sec:appndrres}.

\begin{table}[t]
\begin{center}
\caption{\labell{tab:q7q8RI600and585} Results, at $\beta=6.0$ and 5.85, 
  for the ratios $D_{7,8}^{3/2}$ of \eq{eq:d78def} interpolated to the kaon
  point and our best estimate for $\la (\pi\pi)_{I=2}|Q_{7,8}|K^0\ra$ in the
  chiral limit written as a $K^+\to\pi^+$ matrix element according to
  \eq{eq:kpipiproptokpi}, to avoid ambiguities due to normalization
  conventions for the two pion state.  All results are given in the RI/MOM
  scheme at $2\,\gev$. Please see text for a discussion of how the central
  value and systematic error on the matrix elements is obtained.}
\begin{tabular}{ccccccc}
\hline
\hline
$\beta$ & $D_{7}^{3/2}(M_K^2)$ $[\gev^2]$ & $D_{8}^{3/2}(M_K^2)$ $[\gev^2]$ & 
$\l[\frac{\la \pi^+|Q_{7}^{3/2}|K^+\ra}{F}\r]_\chi$ $[\gev^3]$ & 
$\l[\frac{\la \pi^+|Q_{8}^{3/2}|K^+\ra}{F}\r]_\chi$ $[\gev^3]$\\
\hline
6.0 & $1.63(14)(1)$ & $7.7(11)(0)$ & $0.163(14)(1)(21)$ & $0.77(11)(0)(10)$ \\ 
5.85 & $1.38(34)(21)$ & $6.6(9)(3)$ & $0.138(34)(21)(18)$ & $0.66(9)(3)(9)$ \\ 
\hline
\hline
\end{tabular}
\end{center}
\end{table}

\section{Final physical results and discussion}

\labell{sec:finalres}

As seen the preceding Section, results at $\beta=6.0$ and 5.85 are
in excellent agreement. We could therefore assume that we are in the
scaling regime and fit the $a^2$ dependence of our results to a
constant. However, as discussed briefly in \sec{sec:npr}, the
determination of the renormalization constants is difficult at
$\beta=5.85$.  We therefore choose to take the results obtained on our
finer lattice as our final central values and use the results at
$\beta=5.85$ to estimate possible discretization errors. The lattice
spacing squared at $\beta=6.0$ is essentially half way between that at
$\beta=5.85$ and the continuum limit. Assuming that a linear
extrapolation of the results in $a^2$ to the continuum limit gives a
worst case scenario, we estimate our symmetric discretization errors
to be the absolute value of the difference of the central values of
the results obtained at the two lattice spacings. We then add this
error in quadrature with the error coming from the systematic
uncertainty in the renormalization constants. The procedure assumes
that $O(a^2)$ errors are dominant over higher order discretization
errors and that their coefficient is the same at the two values of
$\beta$, as discussed in \sec{sec:numerics}.

\begin{table}[t]
\begin{center}
\caption{\labell{tab:finalresRImk} Final results at the kaon mass in
  the RI/MOM scheme at $2\,\gev$. The first error is statistical and
  the second is systematic and is described in the text. Also shown
  for comparison are the results of \cite{Donini:1999nn}, obtained using
  quenched, tree-level, $O(a)$-improved Wilson fermions.}
\begin{tabular}{ccccccc}
\hline
\hline
$i$ & \multicolumn{2}{c}{$B_i$} & \multicolumn{2}{c}{$G_i$} & \multicolumn{2}{c}{$R_i$} \\
\hline
& This work & \cite{Donini:1999nn} & This work & \cite{Donini:1999nn} & This work & \cite{Donini:1999nn} \\
\hline
1 & $0.563(47)(30)$ & $0.69(21)$ & $28.8(25)(16)$ & $37.(9)$ & $1$ & $1$ \\ 
2 & $0.865(72)(35)$ & $0.70(9)$ & $-464.(96)(35)$ & 
$-0.24(3)\times 10^3$ & $-16.1(30)(6)$ & $-6.4(18)$ \\ 
3 & $1.41(10)(13)$ & $1.1(1)$ & $151.(30)(7)$ & $74.(9)$ & $5.24(93)(16)$ & $2.0(6)$ \\ 
4 & $0.938(48)(36)$ & $1.1(1)$ & $601.(87)(87)$ & $4.4(4)\times 10^2$ & $20.7(27)(18)$ & $12.(3)$ \\ 
5 & $0.616(51)(59)$ & $0.77(11)$ & $128.(11)(20)$ & $1.0(2)\times 10^2$ & $4.57(60)(68)$ & $2.8(8)$ \\ 
\hline
\hline
\end{tabular}
\end{center}
\end{table}

Let us first consider the BSM matrix elements. We give our final
results for the $B$-parameters, couplings $G_i$ and BSM ratios $R_i$
in \tab{tab:finalresRImk}, in the RI/MOM scheme at $2\,\gev$. In this
table we also provide the results derived from the only other
calculation of these matrix elements \cite{Donini:1999nn}. The latter
were obtained in the quenched approximation, using tree-level,
$O(a)$-improved Wilson fermions. The first point that we wish to make
is that our result for $B_K=B_1$ is in excellent agreement with the
recent world averages of lattice results given in
\cite{Lellouch:2002nj,Hashimoto:2004hn,Dawson:2005za}. The second
point is that our BSM matrix elements are enhanced at $M_K$ by a
factor of five to twenty compared to the chirally suppressed standard
model one, as shown by the values that we obtain for the BSM ratios
$R_i$. The third point is that this enhancement is significantly
larger than the one found in \cite{Donini:1999nn}. This difference can
only partially be explained by our smaller value of $B_1$, which
accounts for about a 23\% enhancement in our results for $R_i$ over
those of \cite{Donini:1999nn}. The remainder of the enhancement must
therefore come from elsewhere. It does not come from the
$B$-parameters $B_i$ for $i=2,\cdots,5$, which agree within errors.
Since the $B$-parameters agree much better than do the matrix elements
themselves, we must already disagree on the vacuum saturation values,
that is on the matrix element of the pseudoscalar density, $\la 0|\bar
s\gamma_5 d|K^0\ra$. Since the axial ward identity requires that $\la
0|\bar s\gamma_5 d|K^0\ra=\sqrt2 F_K M_K^2/(m_s+\hat m)$, we must in
fact find a smaller $(m_s+\hat m)$. Using the results of
\tabs{tab:msplusmhatRI600and585}{tab:msplusmhatMSbar600and585} and the
method described above for obtaining final results on the matrix
elements, we find, for the quark masses:
\be
(m_s+\hat m)^\RI(2\,\gev)=120.(8)(7)\,\mev\qquad\leftrightarrow \qquad
(m_s+\hat m)^\msbar(2\,\gev)=102.(7)(6)\,\mev
\labell{eq:msmhatriandmsbarfinal}
\ ,\ee
which is in excellent agreement with the earlier Neuberger fermion
calculations of \cite{Giusti:2001pk,Babich:2005ay}. Our result for
this sum of quark masses is almost 28\% lower than the one found by
the authors of \cite{Donini:1999nn}, using tree-level, $O(a)$-improved
Wilson fermions at the couplings which they use to obtain the BSM
matrix elements. Indeed, in \cite{Gimenez:1998uv}, they obtain
$(m_s+\hat m)^\msbar(2\,\gev)=130(2)(18)\,\mev$ with the same gauge
configurations as in \cite{Donini:1999nn}. Since this mass appears
squared in the relation of the BSM matrix elements to their respective
$B$-parameters, this difference accounts for the large enhancement of
our BSM matrix elements over those of \cite{Donini:1999nn} and,
combined with our lower value of $B_1$, for the factor of roughly two
enhancement of our BSM ratios $R_i$. Note that the result of
\eq{eq:msmhatriandmsbarfinal} is compatible with the rigorous lower
bounds calculated in \cite{Lellouch:1997hp,Lellouch:1997yn}.

For completeness, we determine the value of the strange quark mass. Using
$m_s/\hat m=24.4\pm 1.5$ as obtained from chiral perturbation theory
\cite{Leutwyler:1996qg}, the results in \eq{eq:msmhatriandmsbarfinal}
correspond to
\be
m_s^\RI(2\,\gev)=116.(8)(7)\,\mev\qquad\leftrightarrow \qquad
m_s^\msbar(2\,\gev)=98.(6)(6)\,\mev
\labell{eq:msriandmsbarfinal}
\ ,\ee
where the uncertainty in the chiral
perturbation theory mass ratio does not contribute significantly to
the error. Our result is fully compatible with the
quenched, continuum limit, benchmark result of the ALPHA collaboration,
$m_s^\msbar(2\,\gev)=97.(4)\,\mev$, obtained using $F_K$, as we do, to
set the lattice spacing \cite{Garden:1999fg}. We therefore believe
that the results of \cite{Donini:1999nn} for the BSM matrix elements
suffer from large systematic errors which are absent in our
calculation.

Because the $\msbar$-NDR scheme is frequently used, we also provide our
results for the $B$-parameters, couplings $G_i$ and BSM ratios $R_i$
in this scheme at $2\,\gev$ in \tab{tab:finalresNDRmk}.

\begin{table}[t]
\begin{center}
\caption{\labell{tab:finalresNDRmk} Final results at the kaon mass in
  the NDR scheme at $2\,\gev$. The first error is statistical and the
  second is systematic and is described in the text.}
\begin{tabular}{cccc}
\hline
\hline
$i$ & $B_i$ & $G_i$ & $R_i$ \\ 
\hline
1 & $0.571(48)(30)$ & $29.2(26)(16)$ & $1$ \\ 
2 & $0.679(56)(27)$ & $-512.(106)(39)$ & $-17.5(32)(7)$ \\ 
3 & $1.055(77)(98)$ & $159.(31)(7)$ & $5.44(97)(15)$ \\ 
4 & $0.810(41)(31)$ & $730(105)(106)\times 10^{3}$ & $24.7(32)(22)$ \\ 
5 & $0.562(39)(46)$ & $165.(15)(25)$ & $5.76(73)(78)$ \\ 
\hline
\hline
\end{tabular}
\end{center}
\end{table}

We turn now to the matrix elements of the electroweak penguin
operators, $\la (\pi\pi)_{I=2}|Q_{7,8}|K^0\ra$. We compute
them in the chiral limit, as discussed in \sec{sec:massdep}. We obtain
final results, as for the BSM matrix elements and quark masses, by
combining our results at $\beta=6.0$ and 5.85 in the manner discussed
above. These results are summarized in \tab{tab:q7q8RIandNDR}, both in the
RI/MOM and $\msbar$-NDR schemes at $2\,\gev$. 

\begin{table}[t]
\begin{center}
\caption{\labell{tab:q7q8RIandNDR} Final results for $\la
  (\pi\pi)_{I=2}|Q_{7,8}|K^0\ra$ in the chiral limit written as a
  $K^+\to\pi^+$ matrix element according to \eq{eq:kpipiproptokpi}, so
  as to avoid ambiguities due to normalization conventions for the two
  pion state.  Results are given in the RI/MOM and $\msbar$-NDR
  schemes at $2\,\gev$. Please see text for a discussion of how the
  central value and systematic error on the matrix elements is
  obtained.}
\begin{tabular}{ccccc}
\hline
\hline
scheme &
$\l[\frac{\la \pi^+|Q_{7}^{3/2}|K^+\ra}{F}\r]_\chi$ $[\gev^3]$ & 
$\l[\frac{\la \pi^+|Q_{8}^{3/2}|K^+\ra}{F}\r]_\chi$ $[\gev^3]$\\
\hline
RI/MOM & $0.163(14)(33)$ & $0.77(11)(15)$ \\
NDR & $0.211(20)(42)$ & $0.93(13)(18)$ \\
\hline
\hline
\end{tabular}
\end{center}
\end{table}

These matrix elements have already been obtained, both in quenched
lattice QCD and with a variety of analytical techniques. In
\tab{tab:q7q8NDRcomp}, we compare our results with those obtained by
others in the NDR scheme at $2\,\gev$, also in the chiral limit. The
results quoted for CP-PACS \cite{Noaki:2001un} were obtained by
multiplying their results for $\la (\pi\pi)_{I=2}|Q_{7,8}|K^0\ra$ at
$1.3\,\gev$ in the $\msbar$-NDR scheme by
$-\sqrt{2/3}(F_\chi+F_K)/(2F_\pi)$, to match our definition of the
matrix elements. They were then run up from 1.3 to $2\,\gev$ using
$N_f=0$ two loop running with the parameters discussed in
\app{sec:appndrrenormcsts}. Those of RBC \cite{Blum:2001xb} were
obtained from the expression $\l[{\la
\pi^+|Q_{7,8}^{3/2}|K^+\ra}/{F}\r]_\chi^{bare}=
(F_\chi+F_K)({b^{8,8}_{(7,8),0}}/{f^2})\times a^{-4}$ with
$f=137(10)\,\mev$ \cite{Blum:2000kn}.  These bare results were then
renormalized in the RI/MOM scheme with the renormalization constants
given in the paper at $2.13\,\gev$, matched onto the $\msbar$-NDR
scheme and finally run down to $2\,\gev$ with our two-loop running
formulae. Finally, the results from SPQcdR \cite{Boucaud:2004aa} are
obtained by extrapolating, to the chiral limit, quenched results for
$K\to\pi\pi$ matrix elements computed for a variety of initial and
final state kinematics, using NLO chiral perturbation theory.

\begin{table}[t]
\begin{center}
\caption{\labell{tab:q7q8NDRcomp} Comparison of our results for the
  matrix elements of the electroweak penguin operators in the chiral
  limit, with those obtained in other lattice calculations (above the
  double line) and with a variety of analytical techniques (below the
  double line). The ``??'' indicate that we have not estimated
  systematic errors in converting the results to our
  conventions. Moreover, the factors of $\sqrt{2/3}$ in front of the
  SPQcdR results \protect\cite{Boucaud:2004aa} are required, we
  believe, to translate their results to our
  normalization.~\protect\footnote{We thank C.-J. David Lin for correspondence
  on this subject.} Results are given in the $\msbar$-NDR scheme at
  $2\,\gev$. }
\begin{tabular}{cccc}
\hline
\hline
Ref. & Action & $\l[\frac{\la \pi^+|Q_{7}^{3/2}|K^+\ra}{F}\r]_\chi$
$[\gev^3]$
& $\l[\frac{\la \pi^+|Q_{8}^{3/2}|K^+\ra}{F}\r]_\chi$ $[\gev^3]$\\
\hline
This work  & Neuberger   & 0.211(20)(42) & 0.93(13)(18)\\
CP-PACS'01 \cite{Noaki:2001un}& Domain-Wall & 0.220(6)(??)    &   0.92(3)(??) \\
RBC'02  \cite{Blum:2001xb} & Domain-Wall & 0.255(12)(??)    &   1.02(4)(??) \\
SPQcdR'04 \cite{Boucaud:2004aa} & Wilson & $\sqrt{2/3}\times 0.16(3)$ & $\sqrt{2/3}\times 0.82(15)$\\
\hline
\hline
Bijnens et al '01 \cite{Bijnens:2001ps}&            & 0.24(3)            & 1.2(7)          \\
Cirigliano et al '02 \cite{Cirigliano:2002jy}&         & 0.22(5)            & 1.50(27)          \\
Friot et al '04 \cite{Friot:2004ba} &              & 0.12(2)            & 2.00(36) \\
Knecht et al '01 \cite{Knecht:2001bc}  &            & 0.11(3)            & 3.5(1.1)          \\
Narison '00  \cite{Narison:2000ys} &            & 0.21(5)            & 1.40(35)          \\
\hline
\hline
\end{tabular}
\end{center}
\end{table}

Agreement of our results with lattice calculations performed using
domain-wall fermions is excellent for both matrix elements. Only the
Wilson fermion results of \cite{Boucaud:2004aa} differ
significantly. This may be due to the absence of chiral symmetry in
the Wilson discretization which significantly complicates the
renormalization procedure and allows potentially large $O(a)$
discretization errors. In regards to the results obtained using
analytical techniques, agreement is found with the results of
\cite{Bijnens:2001ps,Narison:2000ys} and to a lesser extent
\cite{Cirigliano:2002jy}. The authors of
\cite{Knecht:2001bc,Friot:2004ba}, however, find results for the
matrix element of $Q_7^{3/2}$ which are a factor of two lower than
ours, and for the matrix element of $Q_8^{3/2}$, a factor of two to
three higher.

It should be noted that a quenched investigation of electroweak
penguin matrix elements with modified overlap fermions, and
perturbative renormalization, has been performed in
\cite{DeGrand:2003in}. However, the functional form for the chiral
extrapolation of the matrix elements is incorrect and at the kaon
mass, the results obtained at two different lattice spacings do not
appear to be compatible. In addition, preliminary results from a
quenched calculation using HYP staggered fermions were presented in
\cite{Bhattacharya:2004qu}.  Finally, SPQcdR has also attempted to
determine the $\la (\pi\pi)_{I=2}|Q_{7,8}|K^0\ra$ matrix elements at
NLO in chiral perturbation theory by fitting the required low energy
constants to quenched lattice results for a variety of initial and
final state kinematics \cite{Boucaud:2004aa}.

\section{Conclusion}

\labell{sec:ccl}

We have presented results for the full set of $\Delta S=2$ matrix
elements which are required to study $K^0$-$\bar K^0$ mixing in the
standard model and possible extensions. We have also presented
results, at leading chiral order, for the matrix elements of the
electroweak penguin operators $Q_{7,8}^{3/2}$, which give the dominant
$\Delta I=3/2$ contribution to $\epsilon'$. As a by-product of our
calculation, the strange quark mass was obtained. Quarks were
simulated with Neuberger fermions, which possess an exact chiral
flavor symmetry at finite lattice spacing. This not only greatly
simplifies the complicated mixing pattern under renormalization of the
operators that we consider, but also guarantees that our results are
free of the leading $O(a)$ discretization errors. To avoid
potentially large perturbative errors, we implemented all
renormalizations non-perturbatively in the RI/MOM scheme, modifying
the techniques of \cite{Martinelli:1995ty} to accommodate
non-perturbative power corrections and high-momentum discretization
errors. For most renormalization constants, we find that these effects
are important. The calculations were performed on two sets of quenched
configurations, generated with the Wilson gauge action on an
$18^3\times 64$ lattice at $\beta=6.0$ and on a $14^3\times 48$
lattice at $\beta=5.85$. The two lattice spacings allowed us to
investigate scaling violations. Within our statistics, we find no
evidence for such violations and use the results on our coarser
lattice to estimate possible discretization errors.

Our main conclusion is that the non-SM, $\Delta S=2$ matrix elements
are significantly larger than found in the only other dedicated
lattice study of these amplitudes \cite{Donini:1999nn}. In tracing the
source of this difference, we found that we already disagree on
the much simpler matrix element of the pseudoscalar density between a
kaon state and the vacuum, which is the building block for the vacuum
saturation values of the BSM $\Delta S=2$ amplitudes. Through the
axial Ward identity, the matrix element of the pseudoscalar density is related
to the sum of the strange and down quark masses, which we find to be
roughly 30\% smaller than the value obtained in \cite{Gimenez:1998uv},
with the same tree-level improved Wilson fermion action and gauge
configurations as used in \cite{Donini:1999nn}. Since our result for
this sum of masses is in agreement with the continuum limit, benchmark
result of \cite{Garden:1999fg}, we are convinced that the stronger
enhancement of non-SM $\Delta S=2$ matrix elements that we observe is
correct.

Regarding $B_K$ and the matrix elements of the electroweak penguin
operators $Q_{7,8}^{3/2}$, we find good agreement with recent quenched
lattice calculations, in particular those which make use of
domain-wall fermions
\cite{Aoki:1998nr,Aoki:2005ga,Noaki:2001un,Blum:2001xb}, which have an
approximate chiral symmetry at finite lattice spacing. Agreement with
analytical results depends on the particular calculation. 

Beyond this, our investigation validates the use of Ginsparg-Wilson
fermions for the calculation of weak matrix elements. The simplified
operator mixing and non-perturbative $O(a)$ improvement guaranteed by
the full chiral flavor symmetry of these fermions has proven essential
for obtaining reliable results in this delicate calculation. We hope
to have convinced the reader that these benefits outweigh the numerical
overhead that accompanies the use of these fermions. Of course, our
results suffer from the shortcomings of the quenched approximation,
though we have tried to minimize their impact by considering
dimensionless measures of the various matrix elements. Nevertheless,
because of the numerous advantages already discussed, we believe that
the way to remove this approximation is by studying Ginsparg-Wilson
quarks on unquenched backgrounds generated with numerically cheaper
fermion formulations, such as $O(a)$-improved Wilson fermions. Work in
that direction is beginning.

\begin{acknowledgments}
We thank Leonardo Giusti and Massimo Testa for stimulating
discussions. This work is supported in part by US DOE grant
DE-FG02-91ER40676, EU RTN contract HPRN-CT-2002-00311 (EURIDICE), and
EU grant HPMF-CT-2001-01468.  We thank Boston University and NCSA for use
of their supercomputer facilities.
\end{acknowledgments}


\appendix

\clearpage

\section{Results for the pseudoscalar
meson masses and decay constants at $\beta=6.0$ and $5.85$}

\labell{sec:app2ptfns}

In this appendix, we have regrouped the results for the pseudoscalar
meson masses and decay constants, in lattice units, which 
are discussed in \sec{sec:2ptfns}. Please refer to that section for
explanations.

\begin{table}[ht]
\begin{center}
\caption{\labell{tab:600spectlatunits}
Pseudoscalar meson mass and decay constant, in lattice units, 
as a function of bare quark mass at $\beta=6.0$.}
\begin{tabular}{ccccccc}
\hline
\hline
& \multicolumn{3}{c}{$\la P\bar P\ra$} & 
\multicolumn{3}{c}{$\la (P+S) (\bar P-\bar S)\ra$} \\
\hline
$am_{q}$ & $aM$ & $aF$ & $(M/4\pi F)^2$ & $aM$ & $aF$ & $(M/4\pi F)^2$ \\ 
\hline
0.030 & $0.215(5)$ & $0.0514(16)$ & $0.111(9)$ & $0.212(8)$ & $0.0504(14)$ & $0.112(8)$ \\ 
0.040 & $0.244(4)$ & $0.0529(14)$ & $0.135(9)$ & $0.242(6)$ & $0.0521(14)$ & $0.137(7)$ \\ 
0.060 & $0.295(3)$ & $0.0559(13)$ & $0.176(9)$ & $0.295(4)$ & $0.0556(13)$ & $0.178(7)$ \\ 
0.080 & $0.339(3)$ & $0.0590(12)$ & $0.209(9)$ & $0.340(4)$ & $0.0592(13)$ & $0.209(7)$ \\ 
0.100 & $0.379(2)$ & $0.0620(12)$ & $0.236(9)$ & $0.381(3)$ & $0.0628(13)$ & $0.234(8)$ \\ 
\hline
\hline
\end{tabular}
\end{center}
\end{table}

\begin{table}[ht]
\begin{center}
\caption{\labell{tab:585spectlatunits}
Pseudoscalar meson mass and decay constant 
as a function of bare quark mass at $\beta=5.85$ in lattice units.}
\begin{tabular}{ccccccc}
\hline
\hline
& \multicolumn{3}{c}{$\la P\bar P\ra$} & 
\multicolumn{3}{c}{$\la (P+S) (\bar P-\bar S)\ra$} \\
\hline
$am_{q}$ & $aM$ & $aF$ & $(M/4\pi F)^2$ & $aM$ & $aF$ & $(M/4\pi F)^2$ \\ 
\hline
0.030 & $0.254(7)$ & $0.0727(22)$ & $0.077(6)$ & $0.235(14)$ & $0.0713(26)$ & $0.069(12)$ \\ 
0.040 & $0.287(6)$ & $0.0738(19)$ & $0.096(6)$ & $0.275(11)$ & $0.0721(21)$ & $0.092(10)$ \\ 
0.053 & $0.326(5)$ & $0.0755(18)$ & $0.118(6)$ & $0.320(9)$ & $0.0740(19)$ & $0.118(8)$ \\ 
0.080 & $0.395(4)$ & $0.0795(16)$ & $0.156(6)$ & $0.395(6)$ & $0.0789(18)$ & $0.159(6)$ \\ 
0.106 & $0.454(3)$ & $0.0836(15)$ & $0.187(6)$ & $0.456(5)$ & $0.0838(17)$ & $0.188(6)$ \\ 
0.132 & $0.507(3)$ & $0.0876(14)$ & $0.212(6)$ & $0.512(4)$ & $0.0887(16)$ & $0.211(6)$ \\ 
\hline
\hline
\end{tabular}
\end{center}
\end{table}

\clearpage

\section{Results for the bare $B$-parameters, BSM ratios, ``couplings'' $G_i$
and ratios $D_{7,8}^{3/2}$ as a function of bare quark mass at
$\beta=6.0$ and $5.85$}

\labell{sec:app3ptfns}

In this appendix, we have regrouped the results for the bare
$B$-parameters, BSM ratios, ``couplings'' $G_i$ and ratios
$D_{7,8}^{3/2}$ as a function of bare quark mass at $\beta=6.0$ and
$5.85$. These results are discussed in \sec{sec:3ptfns}. Please refer
to that section for explanations.

\begin{table}[ht]
\begin{center}
\caption{\labell{tab:bareB600}
Bare $B$-parameters as a function of bare quark mass at $\beta=6.0$.}
\begin{tabular}{cccccc}
\hline
\hline
$am_{q}$ & $B_1$ & $B_2$ & $B_3$ & $B_4$ & $B_5$ \\
\hline
0.030 & $0.632(63)$ & $0.697(54)$ & $0.803(55)$ & $0.802(50)$ & $0.793(50)$ \\ 
0.040 & $0.651(51)$ & $0.690(44)$ & $0.797(46)$ & $0.855(44)$ & $0.861(45)$ \\ 
0.060 & $0.682(40)$ & $0.693(34)$ & $0.801(35)$ & $0.923(35)$ & $0.962(39)$ \\ 
0.080 & $0.705(35)$ & $0.703(28)$ & $0.810(29)$ & $0.966(28)$ & $1.040(33)$ \\ 
0.100 & $0.727(32)$ & $0.716(24)$ & $0.822(26)$ & $0.996(24)$ & $1.105(30)$ \\ 
\hline
\hline
\end{tabular}
\end{center}
\end{table}

\begin{table}[ht]
\begin{center}
\caption{\labell{tab:bareB585}
Bare $B$-parameters as a function of bare quark mass at $\beta=5.85$.}
\begin{tabular}{cccccc}
\hline
\hline
$am_{q}$ & $B_1$ & $B_2$ & $B_3$ & $B_4$ & $B_5$ \\
\hline
0.030 & $0.588(73)$ & $0.823(49)$ & $0.999(64)$ & $0.640(42)$ & $0.635(44)$ \\ 
0.040 & $0.617(55)$ & $0.794(43)$ & $0.948(53)$ & $0.721(39)$ & $0.725(42)$ \\ 
0.053 & $0.634(43)$ & $0.773(38)$ & $0.910(45)$ & $0.794(36)$ & $0.811(40)$ \\ 
0.080 & $0.661(34)$ & $0.755(30)$ & $0.872(35)$ & $0.887(33)$ & $0.937(37)$ \\ 
0.106 & $0.686(30)$ & $0.753(26)$ & $0.861(29)$ & $0.940(31)$ & $1.025(37)$ \\ 
0.132 & $0.709(28)$ & $0.759(23)$ & $0.862(25)$ & $0.976(30)$ & $1.094(37)$ \\ 
\hline
\hline
\end{tabular}
\end{center}
\end{table}

\begin{table}[ht]
\begin{center}
\caption{\labell{tab:ribarevsam600}
Bare BSM ratios $R_i^\mathrm{BSM}$ as a function of bare quark mass at $\beta=6.0$.}
\begin{tabular}{ccccc}
\hline
\hline
$am_{q}$ & $R_2^\mathrm{BSM}$ & $R_3^\mathrm{BSM}$ & $R_4^\mathrm{BSM}$ & $R_5^\mathrm{BSM}$ \\ 
\hline
0.030 & $-19.6(32)$ & $4.51(72)$ & $27.0(39)$ & $8.9(13)$ \\ 
0.040 & $-16.6(21)$ & $3.84(47)$ & $24.7(29)$ & $8.29(95)$ \\ 
0.060 & $-13.4(12)$ & $3.10(26)$ & $21.4(17)$ & $7.45(61)$ \\ 
0.080 & $-11.59(79)$ & $2.67(17)$ & $19.1(12)$ & $6.86(43)$ \\ 
0.100 & $-10.37(60)$ & $2.38(13)$ & $17.32(88)$ & $6.40(34)$ \\ 
\hline
\hline
\end{tabular}
\end{center}
\end{table}

\begin{table}[ht]
\begin{center}
\caption{\labell{tab:ribarevsam585}
Bare BSM ratios $R_i^\mathrm{BSM}$ as a function of bare quark mass at $\beta=5.85$.}
\begin{tabular}{ccccc}
\hline
\hline
$am_{q}$ & $R_2^\mathrm{BSM}$ & $R_3^\mathrm{BSM}$ & $R_4^\mathrm{BSM}$ & $R_5^\mathrm{BSM}$ \\ 
\hline
0.030 & $-20.7(33)$ & $5.01(78)$ & $19.3(33)$ & $6.4(11)$ \\ 
0.040 & $-16.9(21)$ & $4.05(49)$ & $18.5(24)$ & $6.19(82)$ \\ 
0.053 & $-14.5(14)$ & $3.42(33)$ & $17.9(17)$ & $6.09(60)$ \\ 
0.080 & $-11.67(79)$ & $2.70(19)$ & $16.5(11)$ & $5.80(39)$ \\ 
0.106 & $-10.05(55)$ & $2.30(13)$ & $15.06(82)$ & $5.47(30)$ \\ 
0.132 & $-8.98(41)$ & $2.039(97)$ & $13.85(66)$ & $5.18(25)$ \\ 
\hline
\hline
\end{tabular}
\end{center}
\end{table}

\begin{table}[ht]
\begin{center}
\caption{\labell{tab:gibarevsam600}
Bare ratios $G_i$ as a function of bare quark mass at $\beta=6.0$.}
\begin{tabular}{cccccc}
\hline
\hline
$am_{q}$ & $G_1$ & $G_2$ & $G_3$ & $G_4$ & $G_5$ \\ 
\hline
0.030 & $12.2(19)$ & $-262.(47)$ & $60.(10)$ & $362.(55)$ & $119.(18)$ \\ 
0.040 & $15.3(19)$ & $-229.(32)$ & $52.9(70)$ & $340.(42)$ & $114.(14)$ \\ 
0.060 & $20.9(19)$ & $-194.(19)$ & $44.8(42)$ & $309.(27)$ & $107.5(95)$ \\ 
0.080 & $25.7(20)$ & $-173.(14)$ & $39.9(30)$ & $285.(19)$ & $102.5(70)$ \\ 
0.100 & $29.9(21)$ & $-160.(11)$ & $36.6(24)$ & $266.(15)$ & $98.5(56)$ \\ 
\hline
\hline
\end{tabular}
\end{center}
\end{table}

\begin{table}[ht]
\begin{center}
\caption{\labell{tab:gibarevsam585}
Bare ratios $G_i$ as a function of bare quark mass at $\beta=5.85$.}
\begin{tabular}{cccccc}
\hline
\hline
$am_{q}$ & $G_1$ & $G_2$ & $G_3$ & $G_4$ & $G_5$ \\ 
\hline
0.030 & $9.2(15)$ & $-298.(47)$ & $72.(11)$ & $278.(41)$ & $92.(14)$ \\ 
0.040 & $11.9(15)$ & $-257.(30)$ & $61.4(71)$ & $280.(31)$ & $94.(11)$ \\ 
0.053 & $15.1(14)$ & $-226.(20)$ & $53.3(46)$ & $279.(23)$ & $94.9(84)$ \\ 
0.080 & $20.9(16)$ & $-190.(12)$ & $43.8(28)$ & $267.(16)$ & $94.2(62)$ \\ 
0.106 & $25.9(17)$ & $-169.4(95)$ & $38.7(21)$ & $254.(13)$ & $92.2(53)$ \\ 
0.132 & $30.4(18)$ & $-156.4(80)$ & $35.5(17)$ & $241.(11)$ & $90.2(46)$ \\ 
\hline
\hline
\end{tabular}
\end{center}
\end{table}

\begin{table}[ht]
\begin{center}
\caption{\labell{tab:a2dibarevsam600}
Bare ratios $D_{7,8}^{3/2}$ in lattice units as a function of bare quark mass at $\beta=6.0$
.}
\begin{tabular}{ccc}
\hline
\hline
$am_{q}$ & $a^2D_8^{3/2}$ & $a^2D_7^{3/2}$ \\
\hline
0.030 & $0.96(11)$ & $0.316(34)$ \\ 
0.040 & $0.951(87)$ & $0.319(28)$ \\ 
0.060 & $0.966(61)$ & $0.336(21)$ \\ 
0.080 & $0.992(45)$ & $0.356(16)$ \\ 
0.100 & $1.023(36)$ & $0.378(14)$ \\ 
\hline
\hline
\end{tabular}
\end{center}
\end{table}

\begin{table}[ht]
\begin{center}
\caption{\labell{tab:a2dibarevsam585}
Bare ratios $D_{7,8}^{3/2}$ in lattice units as a function of bare quark mass at $\beta=5.85$.}
\begin{tabular}{ccc}
\hline
\hline
$am_{q}$ & $a^2D_8^{3/2}$ & $a^2D_7^{3/2}$ \\
\hline
0.030 & $1.47(17)$ & $0.487(55)$ \\ 
0.040 & $1.53(13)$ & $0.511(43)$ \\ 
0.053 & $1.59(10)$ & $0.541(35)$ \\
0.080 & $1.690(74)$ & $0.595(27)$ \\ 
0.106 & $1.773(65)$ & $0.644(26)$ \\ 
0.132 & $1.853(62)$ & $0.693(26)$ \\ 
\hline
\hline
\end{tabular}
\end{center}
\end{table}

\clearpage

\section{Tables of renormalization constants in
  the RI/MOM scheme at $\beta=6.0$ and $5.85$}

\labell{sec:apprirenormcsts}

In this appendix, we have regrouped the tables of renormalization
constants discussed in \sec{sec:npr}. Please refer to that section for
explanations.

\begin{table}[ht]
\begin{center}
\caption{\labell{tab:ZOijRI600lr} Mixing matrix for the operators $O_i$ in
the RI/MOM scheme at $2\,\gev$ for $\beta=6.0$.}
\begin{tabular}{cccccc}
\hline
\hline
$i/j$ & 1 & 2 & 3 & 4 & 5 \\
\hline
1 & $2.125(20)$ & $0$ & $0$ & $0$ & $0$ \\ 
2 & $0$ & $1.851(55)$ & $0.022(9)$ & $0$ & $0$ \\ 
3 & $0$ & $-0.141(7)$ & $2.760(33)$ & $0$ & $0$ \\ 
4 & $0$ & $0$ & $0$ & $1.771(43)$ & $-0.468(21)$ \\ 
5 & $0$ & $0$ & $0$ & $-0.168(6)$ & $2.541(18)$ \\ 
\hline
\hline
\end{tabular}
\end{center}
\end{table}

\begin{table}[ht]
\begin{center}
\caption{\labell{tab:ZOijRI585lr}
Mixing matrix for the operators $O_i$ in the RI/MOM 
scheme at $2\,\gev$ for $\beta=5.85$.}
\begin{tabular}{cccccc}
\hline
\hline
$i/j$ & 1 & 2 & 3 & 4 & 5 \\
\hline
1 & $1.750(24)$ & $0$ & $0$ & $0$ & $0$ \\ 
2 & $0$ & $1.872(205)$ & $-0.133(21)$ & $0$ & $0$ \\ 
3 & $0$ & $-0.217(12)$ & $2.220(53)$ & $0$ & $0$ \\ 
4 & $0$ & $0$ & $0$ & $1.939(184)$ & $-0.358(100)$ \\ 
5 & $0$ & $0$ & $0$ & $-0.224(10)$ & $2.174(35)$ \\ 
\hline
\hline
\end{tabular}
\end{center}
\end{table}

\begin{table}[ht]
\caption{\labell{tab:zsri} Results for the renormalization constants
of the local scalar and pseudoscalar densities in the RI/MOM scheme at
$2\,\gev$, $Z_P^\RI(2\,\gev)=Z_S^\RI(2\,\gev)$, for $\beta=6.0$ and
$\beta=5.85$. The results are obtained from fits to an extended range
of $p^2$ as well as to a more restricted range, as described in the
text.}
\begin{tabular}{ccc}
\hline
\hline
$\beta$ & Extended $p^2$-range & Restricted $p^2$-range\\
\hline
6.0 & 1.229(14) & 1.221(13) \\
5.85 & 1.285(34) & 1.194(30)\\
\hline
\end{tabular}
\end{table}

\begin{table}[ht]
\begin{center}
\caption{\labell{tab:ZBijRI600lr}
  Mixing matrix for the $B$-parameters $B_i$ in the RI/MOM scheme
  at $2\,\gev $ and at $\beta=6.0$.}
\begin{tabular}{cccccc}
\hline
\hline
$i/j$ & 1 & 2 & 3 & 4 & 5 \\
\hline
1 & $0.879(8)$ & $0$ & $0$ & $0$ & $0$ \\ 
2 & $0$ & $1.225(26)$ & $-0.074(30)$ & $0$ & $0$ \\ 
3 & $0$ & $0.019(1)$ & $1.828(45)$ & $0$ & $0$ \\ 
4 & $0$ & $0$ & $0$ & $1.173(16)$ & $-0.930(54)$ \\ 
5 & $0$ & $0$ & $0$ & $-0.037(2)$ & $1.683(36)$ \\ 
\hline
\hline
\end{tabular}
\end{center}
\end{table}

\begin{table}[ht]
\begin{center}
\caption{\labell{tab:ZBijRI585lr}
  Mixing matrix for the $B$-parameters $B_i$ in the RI/MOM scheme
  at $2\,\gev $ and at $\beta=5.85$.}
\begin{tabular}{cccccc}
\hline
\hline
$i/j$ & 1 & 2 & 3 & 4 & 5 \\
\hline
1 & $0.840(12)$ & $0$ & $0$ & $0$ & $0$ \\ 
2 & $0$ & $1.133(98)$ & $0.403(53)$ & $0$ & $0$ \\ 
3 & $0$ & $0.026(2)$ & $1.34(10)$ & $0$ & $0$ \\ 
4 & $0$ & $0$ & $0$ & $1.174(82)$ & $-0.650(201)$ \\ 
5 & $0$ & $0$ & $0$ & $-0.045(3)$ & $1.316(87)$ \\ 
\hline
\hline
\end{tabular}
\end{center}
\end{table}

\begin{table}[ht]
\begin{center}
\caption{\labell{tab:ZOijRI600sr}Mixing matrix for the operators $O_i$
  in the RI/MOM scheme at $2\,\gev$ and at $\beta=6.0$. It is
  obtained from a fit to a limited $p^2$ range and is used to
  determine the systematic error associated with our non-perturbative
  renormalization procedure.}
\begin{tabular}{cccccc}
\hline
\hline
$i/j$ & 1 & 2 & 3 & 4 & 5 \\
\hline
1 & $2.151(18)$ & $0$ & $0$ & $0$ & $0$ \\ 
2 & $0$ & $1.806(40)$ & $0.011(19)$ & $0$ & $0$ \\ 
3 & $0$ & $-0.146(12)$ & $2.788(33)$ & $0$ & $0$ \\ 
4 & $0$ & $0$ & $0$ & $1.774(42)$ & $-0.468(20)$ \\ 
5 & $0$ & $0$ & $0$ & $-0.176(10)$ & $2.550(17)$ \\ 
\hline
\hline
\end{tabular}
\end{center}
\end{table}

\begin{table}[ht]
\begin{center}
\caption{\labell{tab:ZOijRI585sr}Same as
  \protect\tab{tab:ZOijRI600sr}, but for $\beta=5.85$.}
\begin{tabular}{cccccc}
\hline
\hline
$i/j$ & 1 & 2 & 3 & 4 & 5 \\
\hline
1 & $1.742(27)$ & $0$ & $0$ & $0$ & $0$ \\ 
2 & $0$ & $1.689(116)$ & $-0.185(73)$ & $0$ & $0$ \\ 
3 & $0$ & $-0.230(37)$ & $2.251(48)$ & $0$ & $0$ \\ 
4 & $0$ & $0$ & $0$ & $1.869(129)$ & $-0.416(64)$ \\ 
5 & $0$ & $0$ & $0$ & $-0.234(31)$ & $2.179(33)$ \\ 
\hline
\hline
\end{tabular}
\end{center}
\end{table}

\begin{table}[ht]
\begin{center}
\caption{\labell{tab:ZBijRI600sr} Same as
  \protect\tab{tab:ZOijRI600sr}, but for the $B$-parameters $B_i$ at
  $\beta=6.0$.}
\begin{tabular}{cccccc}
\hline
\hline
$i/j$ & 1 & 2 & 3 & 4 & 5 \\
\hline
1 & $0.890(8)$ & $0$ & $0$ & $0$ & $0$ \\ 
2 & $0$ & $1.212(18)$ & $-0.036(63)$ & $0$ & $0$ \\ 
3 & $0$ & $0.020(2)$ & $1.870(50)$ & $0$ & $0$ \\ 
4 & $0$ & $0$ & $0$ & $1.190(17)$ & $-0.941(56)$ \\ 
5 & $0$ & $0$ & $0$ & $-0.039(2)$ & $1.711(38)$ \\ 
\hline
\hline
\end{tabular}
\end{center}
\end{table}

\begin{table}[ht]
\begin{center}
\caption{\labell{tab:ZBijRI585sr}Same as
  \protect\tab{tab:ZOijRI600sr}, but for the $B$-parameters $B_i$ at
  $\beta=5.85$.}
\begin{tabular}{cccccc}
\hline
\hline
$i/j$ & 1 & 2 & 3 & 4 & 5 \\
\hline
1 & $0.836(13)$ & $0$ & $0$ & $0$ & $0$ \\ 
2 & $0$ & $1.184(55)$ & $0.647(245)$ & $0$ & $0$ \\ 
3 & $0$ & $0.032(5)$ & $1.579(109)$ & $0$ & $0$ \\ 
4 & $0$ & $0$ & $0$ & $1.311(63)$ & $-0.875(174)$ \\ 
5 & $0$ & $0$ & $0$ & $-0.055(8)$ & $1.529(96)$ \\ 
\hline
\hline
\end{tabular}
\end{center}
\end{table}

\clearpage

\section{Results for the $B$-parameters, BSM ratios, ``couplings'' $G_i$
and ratios $D_{7,8}^{3/2}$, renormalized in the RI/MOM scheme, and as a
function of bare quark mass at $\beta=6.0$ and $5.85$}

\labell{sec:appmassdep}

In this appendix, we have regrouped the results for the
$B$-parameters, BSM ratios, ``couplings'' $G_i$ and ratios
$D_{7,8}^{3/2}$, renormalized in the RI/MOM scheme at $2\,\gev$, and
as a function of bare quark mass at $\beta=6.0$ and $5.85$. These
results are discussed in \sec{sec:massdep}. Please refer to that
section for explanations.

\begin{table}[ht]
\begin{center}
\caption{\labell{tab:birivsam600}
  $B$-parameters in the RI/MOM scheme at $2\,\gev$ as a function of bare quark
  mass at $\beta=6.0$.}
\begin{tabular}{cccccc}
\hline
\hline
$am_{q}$ & $B_1$ & $B_2$ & $B_3$ & $B_4$ & $B_5$ \\
\hline
0.030 & $0.556(56)$ & $0.869(74)$ & $1.42(11)$ & $0.911(59)$ & $0.588(51)$ \\ 
0.040 & $0.572(46)$ & $0.861(62)$ & $1.405(90)$ & $0.971(53)$ & $0.654(51)$ \\ 
0.060 & $0.599(37)$ & $0.865(48)$ & $1.413(73)$ & $1.046(43)$ & $0.761(50)$ \\ 
0.080 & $0.620(32)$ & $0.876(42)$ & $1.429(63)$ & $1.094(35)$ & $0.852(50)$ \\ 
0.100 & $0.639(29)$ & $0.892(38)$ & $1.449(58)$ & $1.127(30)$ & $0.932(51)$ \\ 
\hline
\hline
\end{tabular}
\end{center}
\end{table}

\begin{table}[ht]
\begin{center}
\caption{\labell{tab:birivsam585}
  $B$-parameters in the RI/MOM scheme at $2\,\gev$ as a function of bare quark
  mass at $\beta=5.85$.}
\begin{tabular}{cccccc}
\hline
\hline
$am_{q}$ & $B_1$ & $B_2$ & $B_3$ & $B_4$ & $B_5$ \\
\hline
0.030 & $0.494(62)$ & $0.96(11)$ & $1.67(12)$ & $0.723(67)$ & $0.42(11)$ \\ 
0.040 & $0.518(48)$ & $0.925(96)$ & $1.59(11)$ & $0.814(71)$ & $0.49(13)$ \\ 
0.053 & $0.533(38)$ & $0.900(89)$ & $1.534(99)$ & $0.895(75)$ & $0.55(14)$ \\ 
0.080 & $0.555(31)$ & $0.879(82)$ & $1.476(92)$ & $0.999(82)$ & $0.66(15)$ \\ 
0.106 & $0.576(28)$ & $0.876(80)$ & $1.461(90)$ & $1.058(86)$ & $0.74(16)$ \\ 
0.132 & $0.595(26)$ & $0.883(80)$ & $1.464(89)$ & $1.096(88)$ & $0.81(17)$ \\ 
\hline
\hline
\end{tabular}
\end{center}
\end{table}

\begin{table}[ht]
\begin{center}
\caption{\labell{tab:ririvsam600}
  BSM ratios $R_i^\BSM$ in the RI/MOM scheme at $2\,\gev$ as a function of
  bare quark mass at $\beta=6.0$.}
\begin{tabular}{ccccc}
\hline
\hline
$am_{q}$ & $R_2^\mathrm{BSM}$ & $R_3^\mathrm{BSM}$ & $R_4^\mathrm{BSM}$ & $R_5^\mathrm{BSM}$ \\
\hline
0.030 & $-17.3(28)$ & $5.65(89)$ & $21.8(31)$ & $4.70(70)$ \\ 
0.040 & $-14.7(18)$ & $4.81(58)$ & $19.9(22)$ & $4.47(53)$ \\ 
0.060 & $-11.9(10)$ & $3.89(32)$ & $17.3(13)$ & $4.19(37)$ \\ 
0.080 & $-10.27(71)$ & $3.35(22)$ & $15.39(89)$ & $3.99(30)$ \\ 
0.100 & $-9.19(54)$ & $2.99(16)$ & $13.93(68)$ & $3.84(26)$ \\ 
\hline
\hline
\end{tabular}
\end{center}
\end{table}

\begin{table}[ht]
\begin{center}
\caption{\labell{tab:ririvsam585}
  BSM ratios $R_i^\BSM$ in the RI/MOM scheme at $2\,\gev$ as a function of
  bare quark mass at $\beta=5.85$.}
\begin{tabular}{ccccc}
\hline
\hline
$am_{q}$ & $R_2^\mathrm{BSM}$ & $R_3^\mathrm{BSM}$ & $R_4^\mathrm{BSM}$ & $R_5^\mathrm{BSM}$ \\
\hline
0.030 & $-22.7(43)$ & $7.9(13)$ & $20.5(38)$ & $4.0(13)$ \\ 
0.040 & $-18.6(31)$ & $6.42(80)$ & $19.7(31)$ & $3.9(12)$ \\ 
0.053 & $-16.0(23)$ & $5.44(54)$ & $19.0(26)$ & $3.9(11)$ \\ 
0.080 & $-12.8(16)$ & $4.31(32)$ & $17.5(22)$ & $3.84(98)$ \\ 
0.106 & $-11.0(13)$ & $3.68(23)$ & $16.0(19)$ & $3.72(89)$ \\ 
0.132 & $-9.9(12)$ & $3.27(18)$ & $14.7(17)$ & $3.60(81)$ \\ 
\hline
\hline
\end{tabular}
\end{center}
\end{table}

\begin{table}[ht]
\begin{center}
\caption{\labell{tab:girivsam600}
  $G_i$ ratios in the RI/MOM scheme at $2\,\gev$ as a function of bare quark
  mass at $\beta=6.0$.}
\begin{tabular}{cccccc}
\hline
\hline
$am_{q}$ & $G_1$ & $G_2$ & $G_3$ & $G_4$ & $G_5$ \\
\hline
0.030 & $26.0(41)$ & $-493.(88)$ & $161.(27)$ & $621.(92)$ & $134.(20)$ \\ 
0.040 & $32.6(41)$ & $-431.(61)$ & $141.(19)$ & $584.(70)$ & $131.(16)$ \\ 
0.060 & $44.4(43)$ & $-365.(37)$ & $119.(11)$ & $530.(44)$ & $128.(12)$ \\ 
0.080 & $54.6(45)$ & $-326.(27)$ & $106.3(82)$ & $488.(31)$ & $127.(10)$ \\ 
0.100 & $63.6(47)$ & $-300.(22)$ & $97.6(65)$ & $455.(23)$ & $125.5(89)$ \\ 
\hline
\hline
\end{tabular}
\end{center}
\end{table}

\begin{table}[ht]
\begin{center}
\caption{\labell{tab:girivsam585}
  $G_i$ ratios in the RI/MOM scheme at $2\,\gev$ as a function of bare quark
  mass at $\beta=5.85$.}
\begin{tabular}{cccccc}
\hline
\hline
$am_{q}$ & $G_1$ & $G_2$ & $G_3$ & $G_4$ & $G_5$ \\
\hline
0.030 & $16.0(27)$ & $-574.(112)$ & $200.(31)$ & $519.(90)$ & $101.(31)$ \\ 
0.040 & $20.9(26)$ & $-494.(80)$ & $170.(20)$ & $522.(76)$ & $104.(30)$ \\ 
0.053 & $26.4(26)$ & $-435.(62)$ & $148.(13)$ & $519.(67)$ & $107.(29)$ \\ 
0.080 & $36.6(29)$ & $-365.(46)$ & $122.5(84)$ & $497.(60)$ & $109.(27)$ \\ 
0.106 & $45.3(32)$ & $-326.(40)$ & $108.6(67)$ & $471.(55)$ & $110.(26)$ \\ 
0.132 & $53.2(34)$ & $-301.(36)$ & $99.7(56)$ & $448.(52)$ & $110.(24)$ \\ 
\hline
\hline
\end{tabular}
\end{center}
\end{table}

\begin{table}[ht]
\begin{center}
\caption{\labell{tab:dirivsam600}
  $D_{7,8}^{3/2}$ ratios in the RI/MOM scheme at $2\,\gev$ and in units of
  $\gev^2$ as a function of bare quark mass at $\beta=6.0$.}
\begin{tabular}{ccc}
\hline
\hline
$am_{q}$ & $D_8^{3/2}$ & $D_7^{3/2}$ \\
\hline
0.030 & $7.8(13)$ & $1.67(27)$ \\  
0.040 & $7.7(11)$ & $1.73(25)$ \\ 
0.060 &	$7.80(93)$ & $1.89(23)$ \\
0.080 &	$8.01(82)$ & $2.08(23)$ \\ 
0.100 &	$8.25(78)$ & $2.27(23)$ \\ 
\hline
\hline
\end{tabular}
\end{center}
\end{table}

\begin{table}[ht]
\begin{center}
\caption{\labell{tab:dirivsam585}
  $D_{7,8}^{3/2}$ ratios in the RI/MOM scheme at $2\,\gev$ and in units of
  $\gev^2$ as a function of bare quark mass at $\beta=5.85$.}
\begin{tabular}{ccc}
\hline
\hline
$am_{q}$ & $D_8^{3/2}$ & $D_7^{3/2}$ \\
\hline
0.030 & $6.10(101)$ & $1.18(35)$ \\ 
0.040 &  $6.31(93)$ & $1.25(36)$ \\ 
0.053 &  $6.57(90)$ & $1.35(37)$ \\ 
0.080 &  $6.98(92)$ & $1.53(39)$ \\ 
0.106 &  $7.31(96)$ & $1.70(41)$ \\ 
0.132 &  $7.63(100)$ & $1.87(43)$ \\
\hline
\hline
\end{tabular}
\end{center}
\end{table}

\clearpage

\section{Non-perturbative results for the renormalization constants in
  the $\msbar$-NDR scheme}

\labell{sec:appndrrenormcsts}

Because they may be of use in future calculations, we give
non-perturbative results for the renormalization constants in the
$\msbar$-NDR scheme. These are obtained from the RGI renormalization
constants obtained with the fits described in \sec{sec:npr}, through
the relation
$Z_{ij}^{\NDR}(2\,\gev)=U^{\NDR}_{ik}(2\,\gev)Z^{\RGI}_{kj}$. Here,
$U^{\NDR}_{ik}(p^2)$ describes the running of the renormalization
constants in the $\msbar$-NDR scheme which, again, we implement at two
loops \cite{Buras:2000if} with $N_f=0$ and $\Lambda_\mathrm{QCD}$ from
\cite{Capitani:1998mq}. The matching of the renormalization constants
for the pseudoscalar density $Z_P=Z_S$ is implemented at four loops
\cite{Chetyrkin:1999pq,vanRitbergen:1997va,Vermaseren:1997fq}, with
$N_f=0$ and the same value of $\Lambda_\mathrm{QCD}$. Our results for
all of these renormalization constants are given in
\tabs{tab:ZOijNDR600lr}{tab:ZBijNDR585sr}, including our estimate of the
systematic error associated with the analysis performed in \sec{sec:npr}.

\begin{table}[ht]
\begin{center}
\caption{\labell{tab:ZOijNDR600lr}
Mixing matrix for the operators $O_i$ in the $\msbar$-NDR
scheme at $2\,\gev$ for $\beta=6.0$.}
\begin{tabular}{cccccc}
\hline
\hline
$i/j$ & 1 & 2 & 3 & 4 & 5 \\
\hline
1 & $2.155(21)$ & $0$ & $0$ & $0$ & $0$ \\ 
2 & $0$ & $2.060(61)$ & $0.007(9)$ & $0$ & $0$ \\ 
3 & $0$ & $-0.081(8)$ & $2.834(33)$ & $0$ & $0$ \\ 
4 & $0$ & $0$ & $0$ & $2.152(53)$ & $-0.346(22)$ \\ 
5 & $0$ & $0$ & $0$ & $-0.209(8)$ & $2.494(18)$ \\ 
\hline
\hline
\end{tabular}
\end{center}
\end{table}

\begin{table}[ht]
\begin{center}
\caption{\labell{tab:ZOijNDR585lr}
Mixing matrix for the operators $O_i$ in the $\msbar$-NDR
scheme at $2\,\gev$ for $\beta=5.85$.}
\begin{tabular}{cccccc}
\hline
\hline
$i/j$ & 1 & 2 & 3 & 4 & 5 \\
\hline
1 & $1.774(24)$ & $0$ & $0$ & $0$ & $0$ \\ 
2 & $0$ & $2.080(228)$ & $-0.152(22)$ & $0$ & $0$ \\ 
3 & $0$ & $-0.179(13)$ & $2.280(54)$ & $0$ & $0$ \\ 
4 & $0$ & $0$ & $0$ & $2.356(224)$ & $-0.227(101)$ \\ 
5 & $0$ & $0$ & $0$ & $-0.277(12)$ & $2.128(34)$ \\ 
\hline
\hline
\end{tabular}
\end{center}
\end{table}

\begin{table}[ht]
\caption{\labell{tab:zsmsbar} Results for the renormalization constants
of the local scalar and pseudoscalar densities in the $\msbar$ scheme at
$2\,\gev$, $Z_P^\msbar(2\,\gev)=Z_S^\msbar(2\,\gev)$, for $\beta=6.0$ and
$\beta=5.85$. The results are obtained from fits to an extended range
of $p^2$ as well as to a more restricted range, as described in the
text.}
\begin{tabular}{ccc}
\hline
\hline
$\beta$ & Extended $p^2$-range & Restricted $p^2$-range\\
\hline
6.0 & 1.457(16) & 1.447(16) \\
5.85 & 1.524(40) & 1.416(36)\\
\hline
\end{tabular}
\end{table}

\begin{table}[ht]
\begin{center}
\caption{\labell{tab:ZBijNDR600lr} Mixing matrix for the
  $B$-parameters $B_i$ in the $\msbar$-NDR scheme at $2\,\gev $ and at
  $\beta=6.0$.}
\begin{tabular}{cccccc}
\hline
\hline
$i/j$ & 1 & 2 & 3 & 4 & 5 \\
\hline
1 & $0.891(9)$ & $0$ & $0$ & $0$ & $0$ \\ 
2 & $0$ & $0.970(21)$ & $-0.018(22)$ & $0$ & $0$ \\ 
3 & $0$ & $0.008(7)$ & $1.335(33)$ & $0$ & $0$ \\ 
4 & $0$ & $0$ & $0$ & $1.014(14)$ & $-0.489(37)$ \\ 
5 & $0$ & $0$ & $0$ & $-0.033(1)$ & $1.174(25)$ \\ 
\hline
\hline
\end{tabular}
\end{center}
\end{table}

\begin{table}[ht]
\begin{center}
\caption{\labell{tab:ZBijNDR585lr} Mixing matrix for the
  $B$-parameters $B_i$ in the $\msbar$-NDR scheme at $2\,\gev $ and at
  $\beta=5.85$.}
\begin{tabular}{cccccc}
\hline
\hline
$i/j$ & 1 & 2 & 3 & 4 & 5 \\
\hline
1 & $0.852(12)$ & $0$ & $0$ & $0$ & $0$ \\ 
2 & $0$ & $0.896(78)$ & $0.328(38)$ & $0$ & $0$ \\ 
3 & $0$ & $0.015(1)$ & $0.982(73)$ & $0$ & $0$ \\ 
4 & $0$ & $0$ & $0$ & $1.015(71)$ & $-0.293(141)$ \\ 
5 & $0$ & $0$ & $0$ & $-0.040(3)$ & $0.916(61)$ \\ 
\hline
\hline
\end{tabular}
\end{center}
\end{table}

\begin{table}[ht]
\begin{center}
\caption{\labell{tab:ZOijNDR600sr} Mixing matrix for the
  operators $O_i$ in the $\msbar$-NDR scheme at $2\,\gev$ and at
  $\beta=6.0$. It is obtained from a fit to a limited $p^2$ range
  and are used to determine the systematic error associated with our
  non-perturbative renormalization procedure.}
\begin{tabular}{cccccc}
\hline
\hline
$i/j$ & 1 & 2 & 3 & 4 & 5 \\
\hline
1 & $2.181(19)$ & $0$ & $0$ & $0$ & $0$ \\ 
2 & $0$ & $2.010(44)$ & $-0.004(19)$ & $0$ & $0$ \\ 
3 & $0$ & $-0.085(14)$ & $2.862(34)$ & $0$ & $0$ \\ 
4 & $0$ & $0$ & $0$ & $2.156(51)$ & $-0.345(21)$ \\ 
5 & $0$ & $0$ & $0$ & $-0.219(12)$ & $2.502(17)$ \\ 
\hline
\hline
\end{tabular}
\end{center}
\end{table}

\begin{table}[ht]
\begin{center}
\caption{\labell{tab:ZOijNDR585sr} Same as
  \protect\tab{tab:ZOijNDR600sr} but for $\beta=5.85$. }
\begin{tabular}{cccccc}
\hline
\hline
$i/j$ & 1 & 2 & 3 & 4 & 5 \\
\hline
1 & $1.767(27)$ & $0$ & $0$ & $0$ & $0$ \\ 
2 & $0$ & $1.874(128)$ & $-0.204(76)$ & $0$ & $0$ \\ 
3 & $0$ & $-0.194(41)$ & $2.311(49)$ & $0$ & $0$ \\ 
4 & $0$ & $0$ & $0$ & $2.271(157)$ & $-0.288(70)$ \\ 
5 & $0$ & $0$ & $0$ & $-0.288(37)$ & $2.132(33)$ \\ 
\hline
\hline
\end{tabular}
\end{center}
\end{table}

\begin{table}[ht]
\begin{center}
\caption{\labell{tab:ZBijNDR600sr} Same as
  \protect\tab{tab:ZOijNDR600sr} but for the
  $B$-parameters $B_i$ at
  $\beta=6.0$. }
\begin{tabular}{cccccc}
\hline
\hline
$i/j$ & 1 & 2 & 3 & 4 & 5 \\
\hline
1 & $0.902(8)$ & $0$ & $0$ & $0$ & $0$ \\ 
2 & $0$ & $0.959(14)$ & $0.010(46)$ & $0$ & $0$ \\ 
3 & $0$ & $0.008(1)$ & $1.366(36)$ & $0$ & $0$ \\ 
4 & $0$ & $0$ & $0$ & $1.029(15)$ & $-0.494(39)$ \\ 
5 & $0$ & $0$ & $0$ & $-0.035(2)$ & $1.194(26)$ \\ 
\hline
\hline
\end{tabular}
\end{center}
\end{table}

\begin{table}[ht]
\begin{center}
\caption{\labell{tab:ZBijNDR585sr} Same as
  \protect\tab{tab:ZOijNDR600sr} but for the
  $B$-parameters $B_i$ at
  $\beta=5.85$. }
\begin{tabular}{cccccc}
\hline
\hline
$i/j$ & 1 & 2 & 3 & 4 & 5 \\
\hline
1 & $0.848(13)$ & $0$ & $0$ & $0$ & $0$ \\ 
2 & $0$ & $0.935(43)$ & $0.508(180)$ & $0$ & $0$ \\ 
3 & $0$ & $0.019(4)$ & $1.153(79)$ & $0$ & $0$ \\ 
4 & $0$ & $0$ & $0$ & $1.133(55)$ & $-0.432(125)$ \\ 
5 & $0$ & $0$ & $0$ & $-0.048(7)$ & $1.064(67)$ \\ 
\hline
\hline
\end{tabular}
\end{center}
\end{table}

\clearpage

\section{Results for the $B$-parameters, BSM ratios, ``couplings'' $G_i$
and ratios $D_{7,8}^{3/2}$, renormalized in the
$\msbar$-$\mathrm{NDR}$ scheme, and as a function of bare quark mass
at $\beta=6.0$ and $5.85$}

\labell{sec:appmassdepndr}

Because $\msbar$-NDR is a popular scheme, we give in this appendix
results for the $B$-parameters, BSM ratios, ``couplings'' $G_i$ and
ratios $D_{7,8}^{3/2}$, in that scheme at $2\,\gev$, and as a function
of bare quark mass at $\beta=6.0$ and $5.85$. While we do not use
these results in our paper, they should be useful for comparisons with
future lattice calculations done in the $\msbar$-NDR scheme. The
results are obtained by combining the bare results of
\sec{sec:3ptfns}, which are tabulated in \app{sec:app3ptfns}, and the
renormalization constants obtained in \app{sec:appndrrenormcsts}, using
the extended $p^2$ range.

\begin{table}[ht]
\begin{center}
\caption{\labell{tab:bindrvsam600}
  $B$-parameters in the $\msbar$-NDR scheme at $2\,\gev$ as a function of bare quark
  mass at $\beta=6.0$.}
\begin{tabular}{cccccc}
\hline
\hline
$am_{q}$ & $B_1$ & $B_2$ & $B_3$ & $B_4$ & $B_5$ \\
\hline
0.030 & $0.564(57)$ & $0.682(58)$ & $1.060(80)$ & $0.787(51)$ & $0.539(41)$ \\ 
0.040 & $0.580(46)$ & $0.676(49)$ & $1.051(67)$ & $0.839(46)$ & $0.593(40)$ \\ 
0.060 & $0.608(37)$ & $0.679(38)$ & $1.057(54)$ & $0.904(37)$ & $0.679(38)$ \\ 
0.080 & $0.629(33)$ & $0.688(33)$ & $1.069(47)$ & $0.945(30)$ & $0.749(37)$ \\ 
0.100 & $0.648(30)$ & $0.701(30)$ & $1.084(43)$ & $0.974(26)$ & $0.810(37)$ \\ 
\hline
\hline
\end{tabular}
\end{center}
\end{table}

\begin{table}[ht]
\begin{center}
\caption{\labell{tab:bindrvsam585}
  $B$-parameters in the $\msbar$-NDR scheme at $2\,\gev$ as a function of bare quark
  mass at $\beta=5.85$.}
\begin{tabular}{cccccc}
\hline
\hline
$am_{q}$ & $B_1$ & $B_2$ & $B_3$ & $B_4$ & $B_5$ \\
\hline
0.030 & $0.501(63)$ & $0.752(83)$ & $1.251(90)$ & $0.624(58)$ & $0.395(83)$ \\ 
0.040 & $0.526(48)$ & $0.726(76)$ & $1.192(79)$ & $0.703(61)$ & $0.453(91)$ \\ 
0.053 & $0.540(39)$ & $0.707(70)$ & $1.147(73)$ & $0.773(65)$ & $0.511(99)$ \\ 
0.080 & $0.563(31)$ & $0.690(65)$ & $1.104(67)$ & $0.863(71)$ & $0.60(11)$ \\ 
0.106 & $0.584(28)$ & $0.688(63)$ & $1.093(66)$ & $0.914(74)$ & $0.66(12)$ \\ 
0.132 & $0.604(26)$ & $0.693(63)$ & $1.095(65)$ & $0.946(76)$ & $0.72(12)$ \\ 
\hline
\hline
\end{tabular}
\end{center}
\end{table}

\begin{table}[ht]
\begin{center}
\caption{\labell{tab:rindrvsam600}
  BSM ratios $R_i^\BSM$ in the $\msbar$-NDR scheme at $2\,\gev$ as a function of
  bare quark mass at $\beta=6.0$.}
\begin{tabular}{ccccc}
\hline
\hline
$am_{q}$ & $R_2^\mathrm{BSM}$ & $R_3^\mathrm{BSM}$ & $R_4^\mathrm{BSM}$ & $R_5^\mathrm{BSM}$ \\
\hline
0.030 & $-18.9(31)$ & $5.87(92)$ & $26.1(37)$ & $5.97(86)$ \\ 
0.040 & $-16.0(20)$ & $4.99(60)$ & $23.9(26)$ & $5.63(63)$ \\ 
0.060 & $-12.9(11)$ & $4.03(34)$ & $20.7(16)$ & $5.18(42)$ \\ 
0.080 & $-11.18(77)$ & $3.47(23)$ & $18.4(11)$ & $4.87(33)$ \\ 
0.100 & $-10.00(59)$ & $3.10(17)$ & $16.68(81)$ & $4.63(28)$ \\ 
\hline
\hline
\end{tabular}
\end{center}
\end{table}

\begin{table}[ht]
\begin{center}
\caption{\labell{tab:rindrvsam585}
  BSM ratios $R_i^\BSM$ in the $\msbar$-NDR scheme at $2\,\gev$ as a function of
  bare quark mass at $\beta=5.85$.}
\begin{tabular}{ccccc}
\hline
\hline
$am_{q}$ & $R_2^\mathrm{BSM}$ & $R_3^\mathrm{BSM}$ & $R_4^\mathrm{BSM}$ & $R_5^\mathrm{BSM}$ \\
\hline
0.030 & $-24.7(47)$ & $8.2(13)$ & $24.6(46)$ & $5.2(14)$ \\ 
0.040 & $-20.3(33)$ & $6.66(83)$ & $23.6(37)$ & $5.1(13)$ \\ 
0.053 & $-17.4(25)$ & $5.64(56)$ & $22.8(32)$ & $5.0(11)$ \\ 
0.080 & $-14.0(18)$ & $4.47(33)$ & $21.0(26)$ & $4.9(10)$ \\ 
0.106 & $-12.0(15)$ & $3.82(24)$ & $19.1(23)$ & $4.64(91)$ \\ 
0.132 & $-10.7(13)$ & $3.39(18)$ & $17.6(21)$ & $4.44(83)$ \\ 
\hline
\hline
\end{tabular}
\end{center}
\end{table}

\begin{table}[ht]
\begin{center}
\caption{\labell{tab:gindrvsam600}
  $G_i$ ratios in the $\msbar$-NDR scheme at $2\,\gev$ as a function of bare quark
  mass at $\beta=6.0$.}
\begin{tabular}{cccccc}
\hline
\hline
$am_{q}$ & $G_1$ & $G_2$ & $G_3$ & $G_4$ & $G_5$ \\
\hline
0.030 & $26.3(42)$ & $-544.(97)$ & $169.(28)$ & $754.(111)$ & $172.(25)$ \\ 
0.040 & $33.0(42)$ & $-476.(67)$ & $148.(20)$ & $709.(84)$ & $167.(20)$ \\ 
0.060 & $45.1(44)$ & $-403.(41)$ & $125.(12)$ & $643.(53)$ & $161.(14)$ \\ 
0.080 & $55.3(46)$ & $-360.(30)$ & $111.8(86)$ & $593.(37)$ & $157.(11)$ \\ 
0.100 & $64.5(47)$ & $-332.(24)$ & $102.6(68)$ & $553.(28)$ & $153.3(98)$ \\ 
\hline
\hline
\end{tabular}
\end{center}
\end{table}

\begin{table}[ht]
\begin{center}
\caption{\labell{tab:gindrvsam585}
  $G_i$ ratios in the $\msbar$-NDR scheme at $2\,\gev$ as a function of bare quark
  mass at $\beta=5.85$.}
\begin{tabular}{cccccc}
\hline
\hline
$am_{q}$ & $G_1$ & $G_2$ & $G_3$ & $G_4$ & $G_5$ \\
\hline
0.030 & $16.2(28)$ & $-633.(124)$ & $210.(33)$ & $630.(109)$ & $133.(33)$ \\ 
0.040 & $21.2(27)$ & $-545.(89)$ & $179.(21)$ & $634.(92)$ & $136.(31)$ \\ 
0.053 & $26.8(26)$ & $-480.(68)$ & $156.(14)$ & $631.(82)$ & $139.(30)$ \\ 
0.080 & $37.1(29)$ & $-402.(52)$ & $128.7(89)$ & $604.(73)$ & $140.(28)$ \\ 
0.106 & $45.9(32)$ & $-359.(44)$ & $114.1(70)$ & $572.(67)$ & $139.(26)$ \\ 
0.132 & $54.0(34)$ & $-332.(40)$ & $104.8(59)$ & $544.(63)$ & $137.(25)$ \\ 
\hline
\hline
\end{tabular}
\end{center}
\end{table}

\begin{table}[ht]
\begin{center}
\caption{\labell{tab:dindrvsam600}
  $D_{7,8}^{3/2}$ ratios in the $\msbar$-NDR scheme at $2\,\gev$ and in units of
  $\gev^2$ as a function of bare quark mass at $\beta=6.0$.}
\begin{tabular}{ccc}
\hline
\hline
$am_{q}$ & $D_8^{3/2}$ & $D_7^{3/2}$ \\
\hline
0.030 & $9.4(15)$ & $2.15(34)$ \\ 
0.040 & $9.3(14)$ & $2.20(31)$ \\ 
0.060 & $9.5(11)$ & $2.37(29)$ \\
0.080 & $9.7(10)$ & $2.57(27)$ \\ 
0.100 & $10.02(94)$ & $2.78(28)$ \\
\hline
\hline
\end{tabular}
\end{center}
\end{table}

\begin{table}[ht]
\begin{center}
\caption{\labell{tab:dindrvsam585}
  $D_{7,8}^{3/2}$ ratios in the $\msbar$-NDR scheme at $2\,\gev$ and in units of
  $\gev^2$ as a function of bare quark mass at $\beta=5.85$.}
\begin{tabular}{ccc}
\hline
\hline
$am_{q}$ & $D_8^{3/2}$ & $D_7^{3/2}$ \\
\hline
0.030 & $7.4(12)$ & $1.56(38)$ \\  
0.040 & $7.7(11)$ & $1.65(38)$ \\  
0.053 & $8.0(11)$ & $1.76(39)$ \\ 
0.080 & $8.5(11)$ & $1.96(41)$ \\ 
0.106 & $8.9(12)$ & $2.15(44)$ \\ 
0.132 & $9.3(12)$ & $2.34(45)$ \\ 
\hline
\hline
\end{tabular}
\end{center}
\end{table}

\clearpage

\section{Results at the experimental quark masses at $\beta=6.0$ and
  $5.85$ in the $\msbar$-NDR scheme}

\labell{sec:appndrres}

Because the $\msbar$-NDR scheme is commonly used, we provide here
results in the $\msbar$-NDR scheme at the individual lattice spacings
for the $B$-parameters and other matrix element ratios at the physical
point. These results are used in \sec{sec:finalres} to obtain our
final physical results in that scheme. They are obtained by
multiplying the RI/MOM results obtained in \sec{sec:massdep} by
$U^{\NDR}(2\,\gev)[U^{\RI}(2\,\gev)]^{-1}$. Here, $U^{\NDR}(p^2)$ and
$U^{\RI}(p^2)$ are the matrices which describe the running of the
matrix elements of interest in the $\msbar$-NDR and RI/MOM scheme,
respectively. In the case of the four-quark operators, we implement
them at two loops \cite{Ciuchini:1995cd,Buras:2000if} with $N_f=0$ and
$\Lambda_\mathrm{QCD}$ from \cite{Capitani:1998mq}, as in
\sec{sec:npr}. The running for the pseudoscalar density is implemented
at four loops
\cite{Chetyrkin:1999pq,vanRitbergen:1997va,Vermaseren:1997fq}, with
$N_f=0$ and the same value of $\Lambda_\mathrm{QCD}$. Our results for
all quantities of interest in the $\msbar$-NDR scheme at $2\,\gev$ and
at the two values of the lattice spacing are summarized in
\tab{tab:msplusmhatMSbar600and585}. The
systematic errors quoted in these tables are obtained in that same way
as in \sec{sec:massdep}.

\begin{table}[ht]
\begin{center}
\caption{\labell{tab:msplusmhatMSbar600and585}
$(m_s+\hat m)$ at $2\,\gev$ in the $\msbar$ scheme as
  obtained from our simulations at $\beta=6.0$ and $\beta=5.85$.
The first error is
statistical and the second comes from the systematic uncertainty in the
determination of $Z_S$. }
\begin{tabular}{cc}
\hline
\hline
$\beta$ & $(m_s+\hat m)^\msbar(2\,\gev)$\\
\hline
6.0 & $102.(7)(2)\,\mev$ \\
5.85 & $107.(5)(8)\,\mev$ \\
\hline
\hline
\end{tabular}
\end{center}
\end{table}

\begin{table}[ht]
\begin{center}
\caption{\labell{tab:ndrresatmk600and585}
Results at the kaon mass in the $\msbar$-NDR scheme at $2\,\gev$ for
  $\beta=6.0$ and 5.85. The first error is statistical and the second originates from
  the systematic uncertainty in the renormalization constants.}
\begin{tabular}{ccccccc}
\hline
\hline
& \multicolumn{3}{c}{$\beta=6.0$} & \multicolumn{3}{c}{$\beta=5.85$}\\
\hline
$i$ & $B_i$ & $G_i$ & $R_i$ & $B_i$ & $G_i$ & $R_i$ \\
\hline
1 & $0.571(48)(7)$ & $29.2(26)(4)$ & $1$ & $0.541(37)(4)$ & $27.7(19)(0)$ & $1$ \\ 
2 & $0.679(56)(7)$ & $-512.(106)(12)$ & $-17.5(32)(6)$ & $0.705(71)(33)$ & $-474.(76)(45)$ & $-17.1(28)(15)$ \\ 
3 & $1.055(77)(44)$ & $159.(31)(5)$ & $5.44(97)(9)$ & $1.143(75)(292)$ & $154.(18)(13)$ & $5.56(66)(51)$ \\ 
4 & $0.810(41)(11)$ & $730(105)(8)$ & $24.7(32)(3)$ & $0.781(63)(91)$ & $624.(84)(29)$ & $22.5(32)(8)$ \\ 
5 & $0.562(39)(12)$ & $165.(15)(1)$ & $5.76(73)(3)$ & $0.517(98)(9)$ & $140.(28)(18)$ & $5.0(11)(6)$ \\ 
\hline
\hline
\end{tabular}
\end{center}
\end{table}

\begin{table}[ht]
\begin{center}
\caption{\labell{tab:q7q8NDR600and585} Results, at $\beta=6.0$ and
  5.85, for the ratios $D_{7,8}^{3/2}$ of \eq{eq:d78def} interpolated
  to the kaon point and our best estimate for $\la
  (\pi\pi)_{I=2}|Q_{7,8}|K^0\ra$ in the chiral limit written as a
  $K^+\to\pi^+$ matrix element according to \eq{eq:kpipiproptokpi}, so
  as to avoid ambiguities due to normalization conventions for the two
  pion state.  All results are given in the $\msbar$-NDR scheme at
  $2\,\gev$. Please see text for a discussion of how the central value
  and systematic error on the matrix elements is obtained.}
\begin{tabular}{ccccccc}
\hline
\hline
$\beta$ & $D_{7}^{3/2}(M_K^2)$ $[\gev^2]$ & $D_{8}^{3/2}(M_K^2)$ $[\gev^2]$ & 
$\l[\frac{\la \pi^+|Q_{7}^{3/2}|K^+\ra}{F}\r]_\chi$ $[\gev^3]$ & 
$\l[\frac{\la \pi^+|Q_{8}^{3/2}|K^+\ra}{F}\r]_\chi$ $[\gev^3]$\\
\hline
6.0 & $2.11(20)(1)$ & $9.3(13)(0)$ & $0.211(20)(1)(27)$ & $0.93(13)(0)(12)$ \\ 
5.85 & $1.79(36)(23)$ & $8.0(11)(4)$ & $0.179(36)(23)(23)$ & $0.80(11)(4)(10)$ \\ 
\hline
\hline
\end{tabular}
\end{center}
\end{table}


\clearpage

\bibliography{bsm}

\end{document}